\documentclass[apj]{emulateapj}
\usepackage{graphicx}
\usepackage{natbib}

\shorttitle{Extended Structures of PNe in H$_{2}$}
\shortauthors{X.\ Fang et al.}

\begin{document}

\title{Extended Structures of Planetary Nebulae Detected in H$_{2}$
Emission\footnotemark[$\ast$]} 

\footnotetext[$\ast$]{Based on observations obtained with WIRCam, 
a joint project of CFHT, Taiwan, Korea, Canada, France, at the 
Canada-France-Hawaii Telescope (CFHT) which is operated by the 
National Research Council (NRC) of Canada, the Institut National 
des Sciences de l'Univers of the Centre National de la Recherche 
Scientifique of France, and the University of Hawaii.} 

\author{Xuan Fang$^{1,2}$\footnotemark[$\dagger$], 
Yong Zhang$^{1,3,4}$, 
Sun Kwok$^{1,2}$\footnotemark[$\ddagger$],
Chih-Hao Hsia$^{5}$, 
Wayne Chau$^{4}$,
Gerardo Ramos-Larios$^{6}$, \\
and Mart\'{i}n A. Guerrero$^{7}$} 
\affil{
$^{1}$Laboratory for Space Research, Faculty of Science, The 
University of Hong Kong, Hong Kong, China \\
$^{2}$Department of Earth Sciences, Faculty of Science, The 
University of Hong Kong, Pokfulam Road, Hong Kong, China \\
$^{3}$School of Physics and Astronomy, Sun Yat-Sen University, 
Zhuhai 519082, China\\
$^{4}$Department of Physics, Faculty of Science, The University of 
Hong Kong, Pokfulam Road, Hong Kong, China \\
$^{5}$Space Science Institute, Macau University of Science and 
Technology, Avenida Wai Long, Taipa, Macau, China \\
$^{6}$Instituto de Astronom\'{i}a y Meteorolog\'{i}a, Av. Vallarta 
No. 2602, Col. Arcos Vallarta, CP 44130, Guadalajara, Jalisco, 
Mexico \\
$^{7}$Instituto de Astrof\'\i sica de Andaluc\'\i a (IAA, CSIC), 
Glorieta de la Astronom\'\i a s/n, E-18008 Granada, Spain}

\footnotetext[$\dagger$]{Visiting Astronomer, Key Laboratory of 
Optical Astronomy, National Astronomical Observatories, Chinese 
Academy of Sciences (NAOC), 20A Datun Road, Chaoyang District, 
Beijing 100101, China}

\footnotetext[$\ddagger$]{Visiting Professor, Department of 
Physics and Astronomy, University of British Columbia, Vancouver, 
B.C., Canada.} 

\email{fangx@hku.hk; zhangyong5@mail.sysu.edu.cn; sunkwok@hku.hk}

\begin{abstract}
We present narrow-band near-infrared images of a sample of 11 
Galactic planetary nebulae (PNe) obtained in the H$_{2}$ 
2.122\,$\mu$m and Br$\gamma$ 2.166\,$\mu$m emission lines and the 
$K_{\rm c}$ 2.218\,$\mu$m continuum.  These images were collected 
with the Wide-field Infrared Camera (WIRCam) on the 3.6\,m 
Canada-France-Hawaii Telescope (CFHT); their unprecedented depth 
and wide field of view allow us to find extended nebular structures
in H$_{2}$ emission in several PNe, some of these being the first 
detection.  The nebular morphologies in H$_{2}$ emission are 
studied in analogy with the optical images, and indication on 
stellar wind interactions is discussed.  In particular, the 
complete structure of the highly asymmetric halo in NGC\,6772 is 
witnessed in H$_{2}$, which strongly suggests interaction with the
interstellar medium.  Our sample confirms the general correlation 
between H$_{2}$ emission and the bipolarity of PNe.  The 
knotty/filamentary fine structures of the H$_{2}$ gas are resolved 
in the inner regions of several ring-like PNe, also 
confirming the previous argument that H$_{2}$ emission mostly comes
from knots/clumps embedded within fully ionized material at the 
equatorial regions.  Moreover, the H$_{2}$ image of the 
butterfly-shaped Sh\,1-89, after removal of field stars, clearly
reveals a tilted ring structure at the waist.  These high-quality 
CFHT images justify follow-up detailed morpho-kinematic studies 
that are desired to deduce the true physical structures of a few 
PNe in the sample. 
\end{abstract}


\section{Introduction} 
\label{sec1}

As descendants of the low- and intermediate-mass 
($<$8--10\,$M_{\sun}$) stars and the evolutionary stage right after 
the asymptotic giant branch (AGB), planetary nebulae (PNe) provide 
a fundamental tool to understand the interaction between the 
stellar-synthesized materials and the interstellar medium (ISM). 
However, the mass-loss process of PNe remains a long-standing 
enigma.  The modern view is that PNe are formed through interacting 
stellar winds \citep[ISW,][]{Kwok83} rather than a sudden ejection 
of the envelope of the progenitor AGB stars \citep{KPF78,Kwok82}. 
The mass-loss history possibly has a dependency on the time and 
direction.  Although the ISW model provides a general theoretical 
frame to understand the shaping of PNe, the formation mechanism of 
detailed nebular structures revealed by sensitive observations, 
such as multi-polar and point-symmetric structures, 
\citep[e.g.,][]{Balick87,Lopez98} is still poorly known.

\begin{table*}
\begin{center}
\caption{Target List}
\label{targets}
\begin{tabular}{llcccccl}
\hline\hline
PN Name   & PNG & R.A.$^{\rm a}$ &  Decl.$^{\rm a}$ & Distance$^{\rm b}$ & $z$$^{\rm c}$    & Age   & Morphology/Ref.\\
          &     & (J2000.0)   &     (J2000.0) &  (kpc)   & (kpc)  & (10$^{4}$~yr)& \\
\hline
Hb\,12    & PN\,G111.8$-$02.8 & 23:26:14.81 & $+$58:10:54.65 & 2.26 &0.110 &  \nodata    & Bipolar/[1]\\
NGC\,6445 & PN\,G008.0$+$03.9 & 17:49:15.21 & $-$20:00:34.50 & 1.38 &0.094 &  5.5--6.0   & Bipolar/[2]\\
NGC\,6543 & PN\,G096.4$+$29.9 & 17:58:33.41 & $+$66:37:58.79 & 1.15 &0.573 &  5.8--12.0  & Elliptical$^{\rm d}$/[3],[4]\\
NGC\,6720 & PN\,G063.1$+$13.9 & 18:53:35.08 & $+$33:01:45.03 & 0.92 &0.221 &  2.3--4.6   & Bipolar/Ellipsoidal?$^{\rm e}$/[5],[6],[7]\\
NGC\,6772 & PN\,G033.1$-$06.3 & 19:14:36.37 & $-$02:42:25.04 & 1.31 &0.144 &  3.2--7.3   & Bipolar/[8]\\
NGC\,6781 & PN\,G041.8$-$02.9 & 19:18:28.08 & $+$06:32:19.29 & 0.72 &0.036 & $>$2.9--7.0 & Bipolar/[8],[9]\\
NGC\,6826 & PN\,G083.5$+$12.7 & 19:44:48.16 & $+$50:31:30:33 & 1.40 &0.307 &  \nodata    & Elliptical/[10]\\
NGC\,6894 & PN\,G069.4$-$02.6 & 20:16:23.96 & $+$30:33:53.17 & 1.50 &0.068 &  \nodata    & Round/[11]\\
NGC\,7009 & PN\,G037.7$-$34.5 & 21:04:10.88 & $-$11:21:48.25 & 1.26 &0.713 & $>$3.6--8.9 & Elliptical/[12]\\
NGC\,7048 & PN\,G088.7$-$01.6 & 21:14:15.22 & $+$46:17:17.52 & 1.81 &0.050 &  3.5--4.1   & Bipolar/[13]\\
Sh\,1-89  & PN\,G089.8$-$00.6 & 21:14:07.63 & $+$47:46:22.17 & 1.85 &0.019 &  6.7--8.1   & Bipolar/[14]\\
\hline
\end{tabular}
\begin{description}
\item[$^{\rm a}$] R.A.\ and Decl.\ were all adopted from 
\citet{Kerber03} except NGC\,6445, whose coordinates were adopted 
from \citet{Loup93}. 
\item[$^{\rm b}$] Distances are adopted from \citet{Frew16}.
\item[$^{\rm c}$] Height above the Galactic Plane.
\item[$^{\rm d}$] The elliptical inner core seems to be nested 
inside two bubbles that join at a common waist \citep{BH04}. 
\item[$^{\rm e}$] Still in debate.  See discusion in 
Section~\ref{sec4:1}. 
\end{description}
\tablecomments{[1]-- \citet{KH07}; [2]--\citet{Vazquez04};
[3]--\citet{Balick04}; [4]--\citet{BH04}; [5]--\citet{Guerrero97}; 
[6]--\citet{Kwok08}; [7]--\citet{ODell13}; [8]--\citet{Kast94}; 
[9]--\citet{Phil11}; [10]--\citet{Corradi03}; 
[11]--\citet{Manchado96}; [12]--\citet{Gon03}; [13]--\citet{Kast96}; 
[14]--\citet{Hua97}.}
\end{center}
\end{table*}


The ISW model predicts the presence of AGB halos which are often 
found around PNe \citep[e.g.,][]{Balick92}.  Different from the 
main nebulae that often appear to be highly asymmetric, the halos 
uniformly display spherical morphology, unless there are 
interactions between the PN halos and the ISM.  Fast collimated 
outflows have been thought to be the primary agent for shaping the 
aspheric PNe \citep{ST98}.  The aspherical morphologies, with a 
majority characterized as bipolar or multipolar, are probably 
developed during the short transition from the early post-AGB to 
the PN phase, called the proto-PN phase \citep[e.g.,][]{Sahai11}. 
However, why and when the stellar winds deviate from spherical 
symmetry are still in debate, although hydrodynamic simulations 
have been attempted to reproduce various axisymmetric shapes 
\citep[e.g.,][]{BF02}. 

Since the extended halos surrounding the main nebula often contain 
more mass than the optically bright PNe, detailed mappings of the 
optical nebular shell and the PN halo are essential for 
understanding the mass-loss history and dynamical evolutionary 
process of PNe.  Because of their distances from the central star 
and the extinction due to dust, AGB halos are very faint in the 
visible light (with typical surface brightness about 10$^{-3}$ of 
the main PN shells).  Optical observations have shown that the 
detected nebular masses are significantly lower than theoretical 
predictions; this is the so-called ``missing mass problem'' in PNe 
\citep[e.g.,][]{Kimura12}.  Search for extended structures around 
the main nebulae of PNe provide essential clues to solve this 
problem.  Wide-field infrared (IR) imaging, which mainly traces 
the dust and molecular hydrogen (H$_{2}$) emission, therefore 
provides a powerful tool to study these extended structures.  A 
case study was presented by \citet{Zhang12a}, who discovered a 
very extended ($\sim$40\arcmin\ in diamete) halo around the Helix 
Nebula (NGC\,7293) in the mid-IR emission; this bow-shaped halo 
signifies an interaction between the stellar wind and the ISM and 
and suggests a complicated history of stellar mass loss. 

The 2.122\,$\mu$m 1$-$0 S(1) ro-vibrational emission line of 
H$_{2}$ is usually bright in warm, dense, molecular regions, and 
has been a commonly-used tracer of molecular gas in a variety of 
astrophysical environments, including PNe.  This H$_{2}$ emission 
line has been detected in a number of Galactic PNe 
\citep[e.g.,][]{Kast94,Kast96,Latter95,Shupe95,Schild95,
Guerrero00,Arias01,Ramos08,Ramos12,Ramos17,Marquez13,Marquez15} 
as well as proto-PNe \citep[e.g.,][]{Sahai98,Hrivnak08,FG12}. 
Imaging studies of the H$_{2}$ images of PNe have revealed that 
the H$_{2}$ 2.122\,$\mu$m emission in PNe is generally associated 
with the bipolar morphology \citep[e.g.,][]{ZG88,Webster88}; this 
is known as the {\it Gatley's rule}, and the possible astrophysics
as to this association is still discussed today 
\citep[e.g.,][]{Ramos17}.  The UKIRT Wide Field Infrared Survey 
for H$_{2}$ (UWISH2\footnote{http://astro.kent.ac.uk/uwish2}; 
\citealt{Froebrich11}) has detected H$_{2}$ emission from $\sim$280
candidate PNe and proto-PNe in the northern Galactic Plane 
\citep{Froebrich15,GF17}, the majority of which were new 
detections, thus providing a rich database for the study of H$_{2}$
emission and PN morphologies. 

H$_{2}$ emission can be excited by fluorescence or radiative 
pumping by the UV radiation from PN central stars 
\citep[e.g.,][]{Bv87,Diner88} or by shocked gas 
\citep[e.g.,][]{SB82,Burton92}.  If the shocks are the dominant 
excitation mechanism, H$_{2}$ line emission can provide significant
probes of the zones of dynamical interaction.  For example, the 
multiple layers of halo in NGC\,6720 have been observed in the 
H$_{2}$ 2.122\,$\mu$m emission, which might be shock excited 
\citep[e.g.,][]{Kast94}.  Moreover, the H$_{2}$ to Br$\gamma$ 
emission line ratio has been found to correlate with the detailed 
morphologies of bipolar PNe \citep[][]{Guerrero00,Ramos17}, which 
might be related to the dominant excitation mechanism of H$_{2}$. 
the Relative strengths of H$_{2}$ emission lines with different 
excitation energies can also be used to trace the evolutionary 
status of PNe \citep[][]{Davis03}. 

\begin{figure*}
\begin{center}
\includegraphics[width=17.8cm,angle=0]{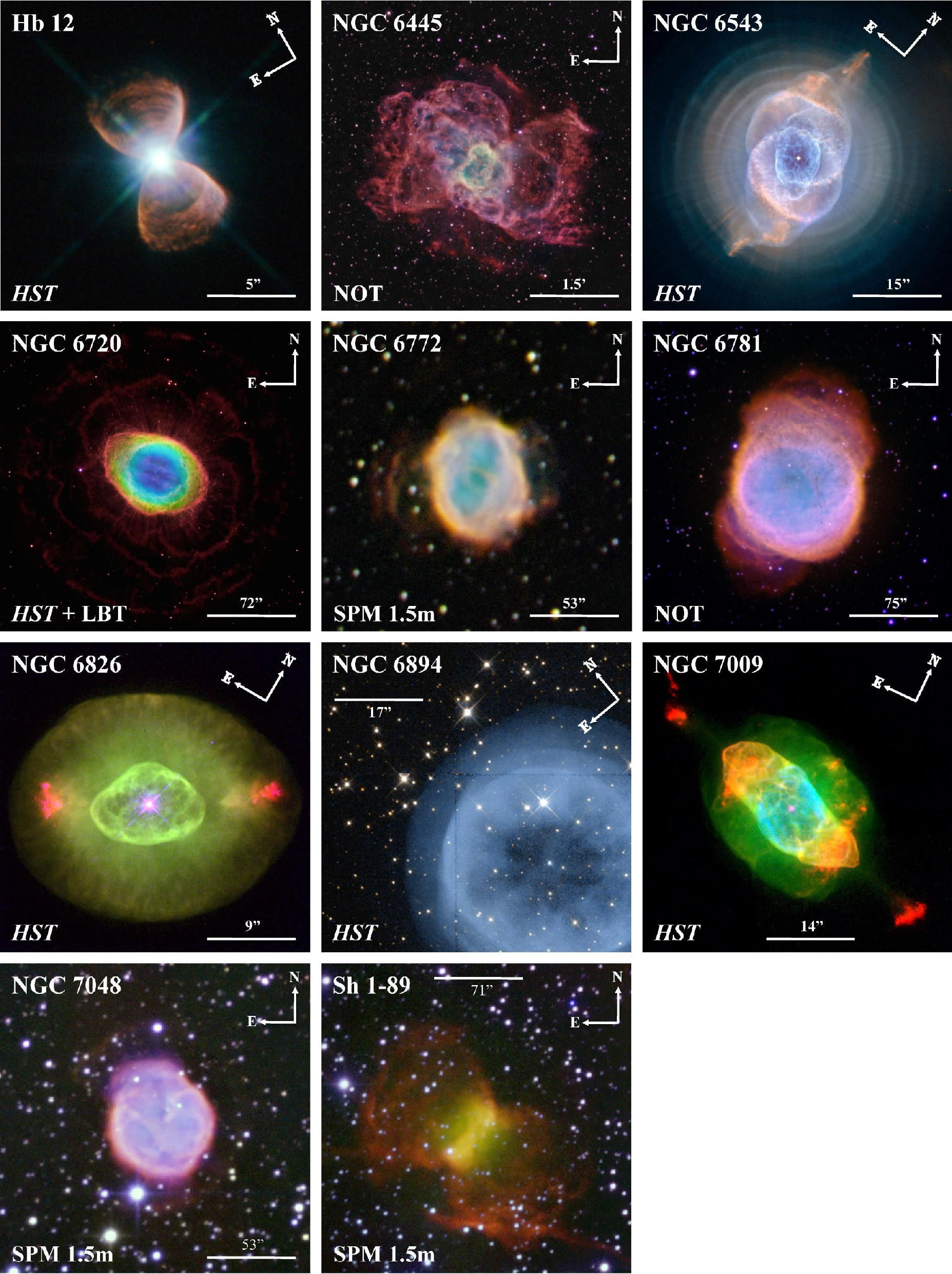}
\caption{Color-composite optical images of our PN sample created 
with the [N~{\sc ii}] (red), H$\alpha$ (green), and [O~{\sc iii}] 
(blue) images, except NGC\,6894, which is combined from the 
\emph{HST} WFC3 F814W (red) and F555W (blue) images.  The 
telescopes used to obtain these images are indicated at the 
bottom-left corner of each panel.  Image scales and orientations 
are marked.} 
\label{fig1}
\end{center}
\end{figure*}

In order to search for extended nebular structures, we obtained 
deep near-IR images of a sample of Galactic PNe (see 
Figure~\ref{fig1}) using the H$_{2}$, Br$\gamma$ and $K_{c}$ 
narrow-band filters on CFHT.  We introduce observations and data 
reduction in Section~\ref{sec2}, and describe morphologies in both 
the near-IR and optical emission lines in Section~\ref{sec3}.  We 
present extensive discussion in Section~\ref{sec4}, and summarize 
our discussion in Section~\ref{sec5}.

\section{Observations and Data Reduction} 
\label{sec2}

\subsection{CFHT Near-IR Observations} 
\label{sec2:1}

The deep near-IR imagery of a dozen Galactic PNe was carried out 
using the Wide-field InfraRed Camera \citep[WIRCam,][]{Puget04} 
on the 3.6\,m Canada-France-Hawaii Telescope (CFHT) in 2012 August 
(program~ID: 12AS98).  The wide-field imager WIRCam is composed 
of four HAWAII2-RG detectors (each 2048$\times$2048 pixels) and 
covers a total field of view (FoV) of 20\arcmin$\times$20\arcmin\ 
with a pixel scale of 0\farcs3.  The background emission from the 
sky and the telescope were removed by chopping and nodding during 
the observations.  The nebular images were obtained in three 
narrow-band filters: H$_{2}$ ($\lambda_{\rm c}$ = 2.122\,$\mu$m, 
$\Delta\lambda$ = 0.032\,$\mu$m), Br$\gamma$ ($\lambda_{\rm c}$ = 
2.166\,$\mu$m, $\Delta\lambda$ = 0.030\,$\mu$m), and $K_{\rm c}$ 
($\lambda_{\rm c}$ = 2.218\,$\mu$m, $\Delta\lambda$ = 
0.033\,$\mu$m.  For each PN, at least ten $\sim$60\,s exposures 
were obtained with each of the three filters. 

Observing nights were photometric; during the whole observing run, 
the seeing varied between 0\farcs5 and 0\farcs7.  The average full 
widths at half-maximum (FWHM) of the point sources in the H$_{2}$, 
Br$\gamma$, and $K_{\rm c}$ filters were 0\farcs55, 0\farcs51, and 
0\farcs56, respectively.  Images with the best quality were 
selected and reduced following standard procedure using 
{\sc iraf}\footnote{{\sc iraf}, the Image Reduction and Analysis 
Facility, is distributed by the National Optical Astronomy 
Observatory, which is operated by the Association of Universities 
for Research in Astronomy under cooperative agreement with the 
National Science Foundation.} packages, including dark subtraction,
flat-field correction, basic crosstalk removal, and sky subtraction
using neighboring images.  Separate exposure frames with sky 
subtraction were combined and mosaicked to produce the final 
images.  A summary of the CFHT observations is given in 
Table~\ref{targets}.  Flux calibration was not performed for our 
targets because no photometric standard stars were observed.

\subsection{Narrow-band Optical Imagery}
\label{sec2:2}

Optical narrow-band images of the PNe NGC\,6445 and NGC\,6781 were 
obtained with the Andaluc\'\i a Faint Object Spectrograph and 
Camera (ALFOSC) on the 2.5\,m Nordic Optical Telescope (NOT) at the 
Observatorio del Roque de los Muchachos (ORM, La Palma, Spain) in 
2009 June.  The optical images of NGC\,6772, NGC\,7048 and Sh\,1-89 
were obtained at the 1.5\,m telescope at San Pedro M\'{a}rtir (SPM) 
of the National Astronomical Observatory (OAN-SPM, Mexico) in 2009 
August and 2010 June.  For NGC\,6445, NGC\,6772, NGC\,6781, and 
NGC\,7048, narrow-band filters centered at the [O~{\sc iii}] 
$\lambda$5007, [N~{\sc ii}] $\lambda$6583 and H$\alpha$ emission 
lines were used in the observations; for Sh\,1-89, the [N~{\sc ii}] 
and H$\alpha$ filters were used.  The images obtained with the 
2k$\times$2k CCD at the NOT have a plate scale of $\sim$0\farcs184 
pixel$^{-1}$, while the 501$\times$501 binned CCD at the 1.5\,m SPM
has a plate scale of 0\farcs504 pixel$^{-1}$.  These optical images
were first registered to the same orientation and sky coordinates 
(in J2000.0), and then rebinned to the same pixel scales as our 
CFHT images.  This will allow a comparison study of nebular 
morphologies in the optical and IR bands.  The color-composite 
images are presented in Figure~\ref{fig1}.

\subsection{The HST Archive Data}
\label{sec2:3} 

Archival \emph{Hubble Space Telescope} (\emph{HST}) optical and 
near-IR narrow-band images were also used in our morphological 
study. These images were retrieved from the Mikulski Archive for 
Space Telescope (MAST\footnote{http://archive.stsci.edu}) at the 
Space Telescope Science Institute.  The \emph{HST} optical images 
were obtained with the Wide-Field Planetary Camera 2 (WFPC2) and 
the Wide Field Camera~3 (WFC3) in conjunction with the F502N, F656N
and F658N narrow-band filters.  The unparalleled high-spatial 
resolution of the WFPC2 (0\farcs0396 pixel$^{-1}$) and WFC3 
(0\farcs040 pixel$^{-1}$) cameras enables detailed inspection of 
the micro-structures of our targets, especially for the inner 
nebular regions, where the optical emission usually dominates. 
The \emph{HST} color-composite images of our targets are presented 
in Figure~\ref{fig1}. 

In particular, for Hb\,12 we also retrieved the WFC3 broad-band 
F140W ($\lambda_{\rm c}$=1.392\,$\mu$m, $\Delta\lambda$=3840\,{\AA})
near-IR images (obtained through the \emph{HST} program~11552, PI: 
H.~A. Bushouse) from the archive.  In the near-IR channel, the WFC3
detector has a FoV of 136\arcsec$\times$136\arcsec\ with a pixel 
scale of 0\farcs13.  The nine F140W images of Hb\,12 (an exposure 
of 2.93\,s each) were first aligned according to the central star 
position, and then combined to enhance the spatial sample and 
remove cosmic rays. 

\subsection{The Spitzer Archive Data}
\label{sec2:4}

Where available, the archival \emph{Spitzer Space Telescope} mid-IR 
images were also used for a comparison study of the PN morphologies.
The \emph{Spitzer} images were observed with the Infrared Array 
Camera \citep[IRAC,][]{Fazio04} employing four filters: 3.6\,$\mu$m 
($\lambda_{\rm c}$ = 3.550\,$\mu$m, $\Delta\lambda$ = 0.75\,$\mu$m),
4.5\,$\mu$m ($\lambda_{\rm c}$ = 4.493\,$\mu$m, $\Delta\lambda$ = 
1.901\,$\mu$m), 5.8\,$\mu$m ($\lambda_{\rm c}$ = 5.731\,$\mu$m, 
$\Delta\lambda$ = 1.425\,$\mu$m), and 8\,$\mu$m ($\lambda_{\rm c}$ 
= 7.872\,$\mu$m, $\Delta\lambda$ = 2.905\,$\mu$m).  The IRAC images 
have plate scales of $\sim$1\farcs6--1\farcs8 pixel$^{-1}$ in these 
four bands, and have a field of view of 5\farcm2$\times$5\farcm2. 
Images were downloaded from the NASA/IPAC Infrared Science Archive 
in the form of \emph{Spitzer} Enhanced Imaging Products 
(SEIP)\footnote{http://irsa.ipac.caltech.edu/data/SPITZER/Enhanced/SEIP/}.

\subsection{The WISE Images}
\label{sec2:5}

In order to facilitate analysis of our sample, we downloaded 
near- and mid-IR images of the \emph{Wide-field Infrared Survey 
Explorer} (\emph{WISE}; \citealt{wright10}) mission from the 
NASA/IPAC Infrared Science Archive 
(IRSA).\footnote{http://irsa.ipac.caltech.edu/applications/wise} 
\emph{WISE} performs an all-sky survey at 3.4\,$\mu$m ($W$1), 
4.6\,$\mu$m ($W$2), 12\,$\mu$m ($W$3), and 22\,$\mu$m ($W$4) with 
angle resolutions of 6\farcs1, 6\farcs4, 6\farcs5, and 12\farcs0, 
respectively.  The \emph{WISE} images in the former three bands 
are used in this work.

\begin{figure*}
\begin{center}
\includegraphics[width=17.7cm,angle=0]{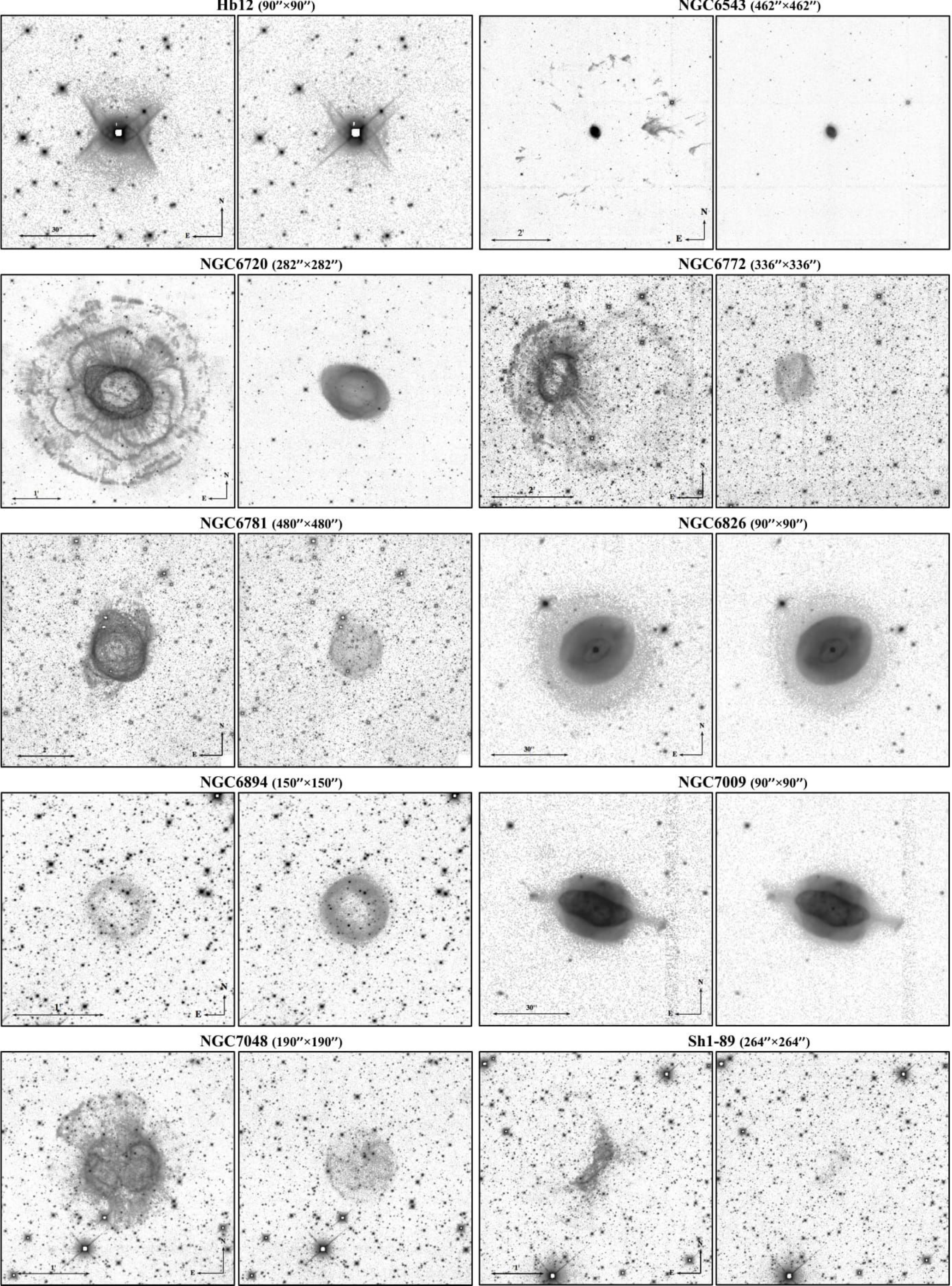}
\caption{CFHT WIRCam negative greyscale images of our sample PNe. 
For each PN, the H$_{2}$ (left) and Br$\gamma$ (right) images are 
displayed; target name and image size (in arcsec) are indicated.} 
\label{fig2}
\end{center}
\end{figure*}

\section{Results}
\label{sec3} 

The CFHT near-IR images of our sample PNe are displayed in 
Figure~\ref{fig2}.  The images of NGC\,6445 are haunted by very 
crowded stellar field, and thus are not presented in greyscale as 
in Figure~\ref{fig2}, but will be shown later on as a 
color-composite picture in this paper.  Some H$_{2}$ images show 
extended halo structures surrounding the central nebulae.  Some of 
these structures generally trace the optical images, but others 
are detected for the first time.  However, the WIRCam $K_{\rm c}$ 
($\lambda_{\rm c}$ = 2.218\,$\mu$m, $\Delta\lambda$ = 0.033\,$\mu$m)
filter used in these observations also includes the 1$-$0 S(0) 
2.223\,$\mu$m and 2$-$1 S(1) 2.243\,$\mu$m line emission of H$_{2}$;
thus strictly speaking, it is not a true $K$-continuum filter that 
can used to subtract from the H$_{2}$ 2.122\,$\mu$m image, as 
usually done in a conventional manner.  That is why in the CFHT 
images of some of our targets, we also see $K_{\rm c}$ emission in 
the nebular halos where the H$_{2}$ emission was detected.  The 
$K_{\rm c}$ images are thus not analyzed in this paper; neither 
are they displayed in Figure~\ref{fig2}.  In this section, we 
compare the CFHT images with the optical images obtained with the 
\emph{HST} and the ground-based telescopes.  Implications for 
wind-wind interaction as well as possible interactions between PNe 
and the ISM are discussed in Section~\ref{sec4}.

\subsection{Hb\,12}
\label{sec3:1}

Hb\,12 was has been classified as a PN by its emission line 
spectrum \citep[e.g.,][]{HA96}. Its compactness as well as high 
surface brightness in the IR \citep{ZK90} and radio \citep{AK90} 
emission indicates that it is a PN at an early stage of evolution, 
which is consistent with a very young kinematic age derived 
through high resolution spectroscopy \citep{MS89}.  Hb\,12 has a 
well-defined hourglass shape in the \emph{HST} optical narrow-band 
images \citep{KH07}.  Spectroscopic observations in the H$_{2}$ 
emission indicate that a huge range of gas density 
(10$^{4}$--10$^{5}$ cm$^{-3}$) exists in this compact object 
\citep{Ramsay93}.  Optical spectroscopy also reveals electron 
density as high as $\sim$500\,000 cm$^{-3}$ \citep{HA96}.  
Fluorescent molecular hydrogen emission was detected in the 
$<$4\arcsec\ inner region of Hb\,12 \citep{Diner88}.  All these 
studies point to the peculiarity of this object. 

\begin{figure*}
\begin{center}
\includegraphics[width=14.0cm,angle=0]{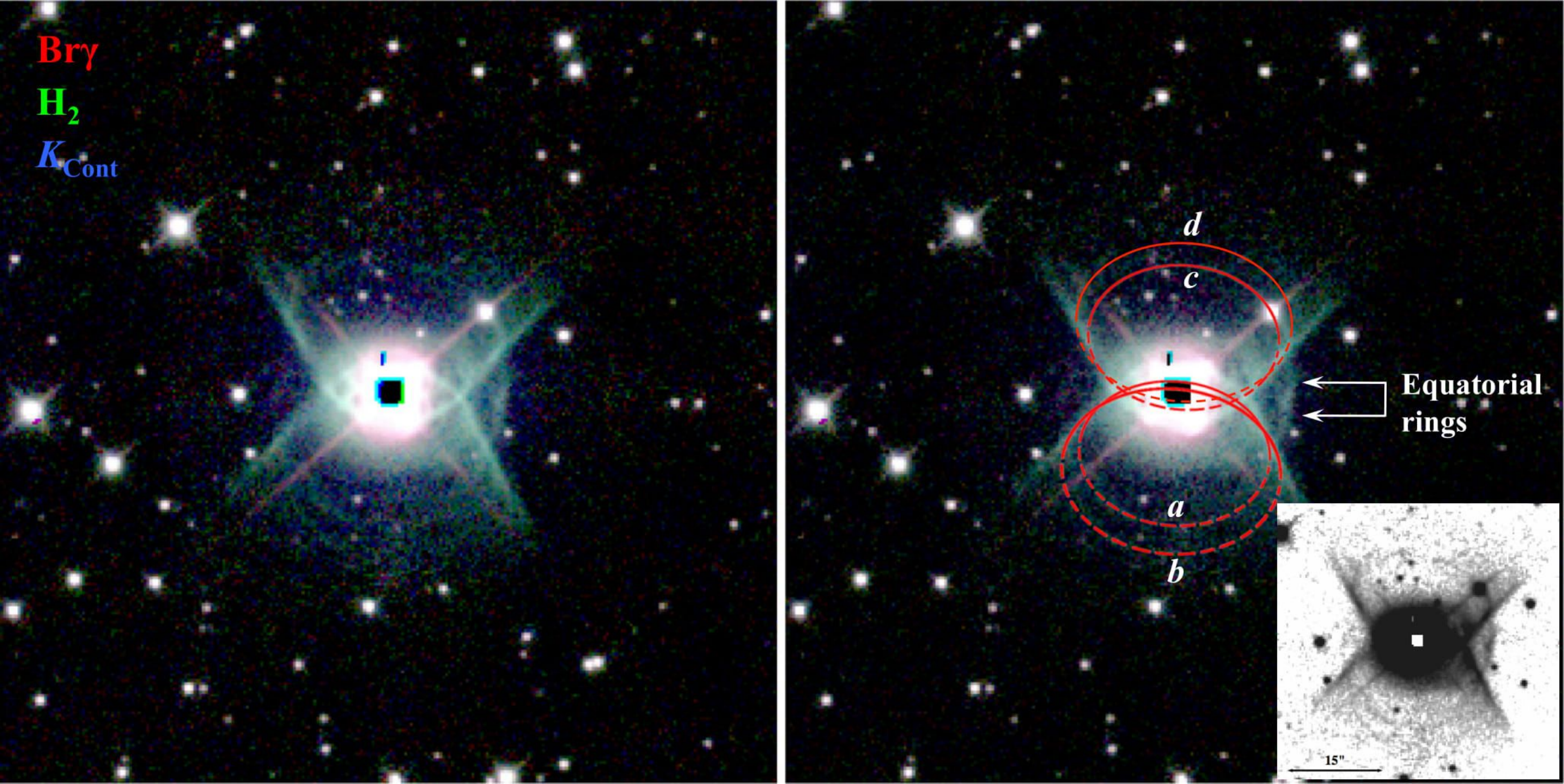}
\caption{Left: CFHT color-composite image of Hb\,12 created with
the Br$\gamma$ (red), H$_{2}$ (green), and $K_{\rm c}$ (blue) 
filters.  Image size is similar to Figure~\ref{fig2}. North is up 
and east to the left.  Right: same as left but over-plotted with 
schematic outlines of four co-axial rings (red ellipses); the front
side of the red ellipses is solid and the back side is dashed. 
Inset shows the H$_{2}$ image scaled to show the rings.}
\label{fig3}
\end{center}
\end{figure*}

Our CFHT images not only show bipolar lobes of Hb\,12, but also 
reveal more detailed structures in H$_{2}$ emission.  This is better
seen in the color-composite image (Figure~\ref{fig3}), where some 
arcs are seen aligned in the bipolar lobes.  A close inspection of 
the image shows that there are two arcs in the Southern lobe, while 
in the Northern lobe, there also seems to be two arcs marginally 
visible.  They are probably due to co-axial, concentric rings on 
either side of the central star, aligned along the bipolar lobes. 
These rings can generally be fitted by ellipses with a 
minor-to-major axes ratio of $\sim$0.7.  If we assume that the 
rings are circular, this ratio indicates a inclination angle (i.e.,
the angle of the polar axis with respect to the plane of the sky) 
of 45$\degr\pm$7$\degr$. 

The two northern rings are about 0.106 and 0.119~pc, or $\sim$21860
and 24500~AU, respectively, from the central star.  The two 
southern rings are 0.110 and 0.133~pc, corresponding to 22750 and 
27420~AU, respectively, from the central star.  These distances of 
the rings have been corrected for the inclination angle (45$\degr$)
of the polar axis, and a distance of 2.26~kpc \citep{Cahn92,Frew16}
to Hb\,12 was assumed.  The opening angle of the bipolar lobes in 
the H$_{2}$ emission is about 75$\degr$. 

\begin{figure*}
\begin{center}
\includegraphics[width=14.0cm,angle=0]{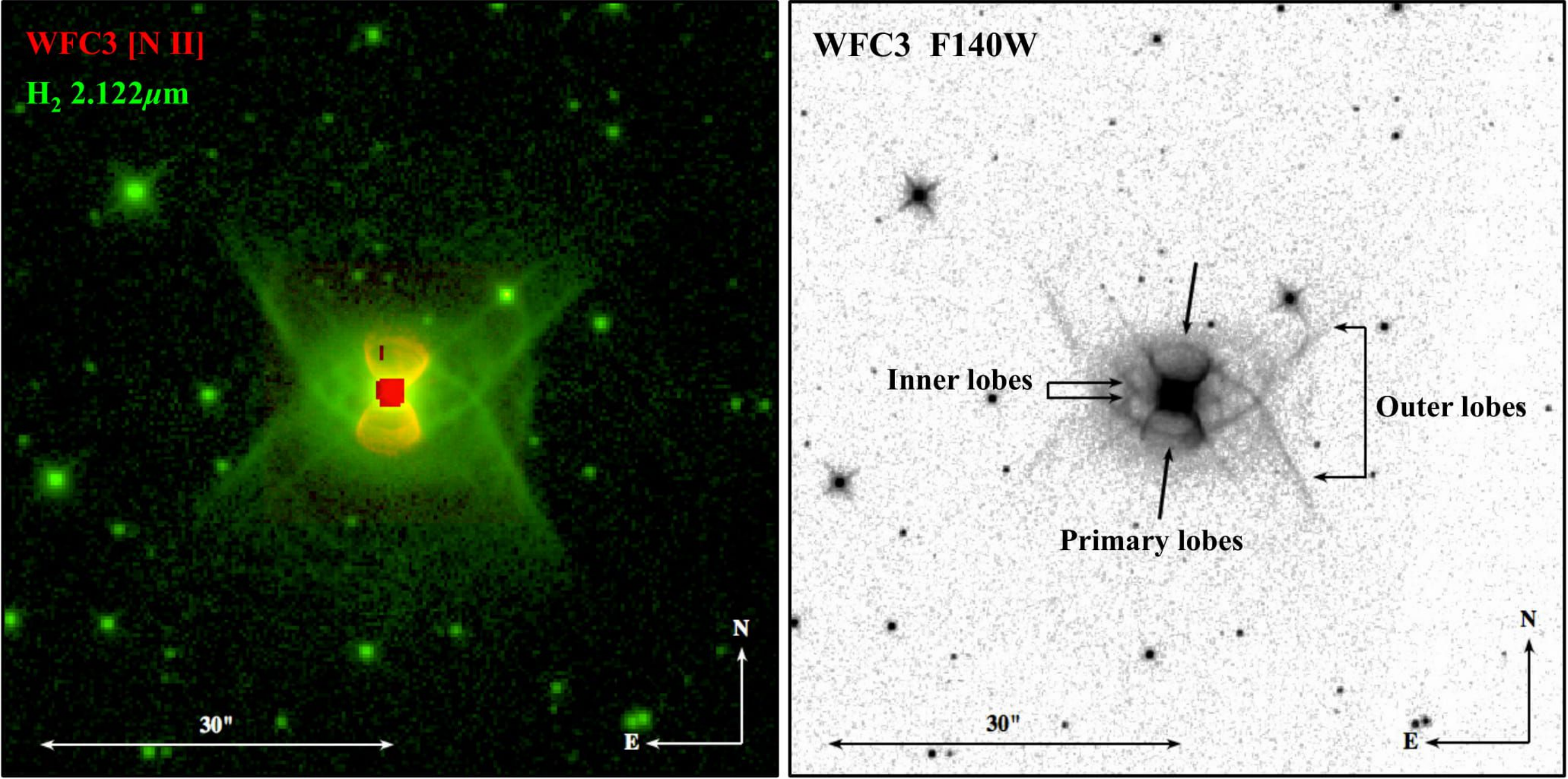}
\caption{Left: color-composite image of Hb\,12 created with the 
\emph{HST} WFC3 F658N (red) and CFHT WIRCam H$_{2}$ (green) images.
Right: \emph{HST} WFC3 F140W image displayed in negative greyscale 
(\emph{HST} program \#11552, PI: H.~A.\ Bushouse). 
The primary bipolar lobes of Hb\,12 defined in the F140W image 
follows the [N~{\sc ii}] morphology (left).  Central arcs in the 
F140W image, probably due to projection of the inner and outer 
bipolar lobes, are also seen in the H$_{2}$ image.} 
\label{fig4}
\end{center}
\end{figure*}

\citet{KH07} proposed that the bipolar lobes Hb\,12 (observed by 
\emph{HST} WFPC2) may represent an inner hourglass nested in a 
much larger bipolar structure, a morphology that is similar to 
some other hourglass-shaped bipolar nebulae such as the 
``Minkowski's Butterfly Nebula'' M\,2-9 \citep[e.g.,][]{Castro12,
Castro17}, Hen\,2-104 \citep{Corradi01}, MyCn\,18 \citep{Bryce97,
Sahai99}, and SN\,1987A \citep{Panagia96,Sugerman05}.  Comparison 
of our CFHT images with the archival \emph{HST} optical image 
indicates that Hb\,12 does exhibit multiple bipolar lobes 
(Figure~\ref{fig4}, left).  Based on the fits of the multiple 
elliptical rings observed in the \emph{HST} WFPC2 [N~{\sc ii}] 
image, \citet{KH07} derived an inclination angle of $\sim$38$\degr$
for the polar axis of Hb\,12.  This inclination angle is consistent
with both the value of 40$\degr$ derived by \citet{HA96} and ours 
(45$\degr\pm$7$\degr$) derived from the H$_{2}$ near-IR image. 
Thus the two pairs of bipolar lobes are aligned with each other and
the southern lobes are tilted towards the observer \citep{HA96}.  
The position angle (PA) of axis of Hb\,12's bipolar lobes is 
171$\fdg5\pm$2$\fdg$3 \citep{KH07}.  The arcs at the equatorial 
region that represent the outer bipolar lobes of Hb\,12 
(Figure~\ref{fig4}, right) was also noticed in the \emph{HST} NICMOS
F212N image \citet{Clark14}. 

Between the outer bipolar lobes observed via H$_{2}$ emission and 
the primary lobes in \emph{HST} F658N (Figure~\ref{fig4}, left), 
there seems to be a third pair of bipolar lobes, which can be 
inferred from a smaller eye-shaped feature shown in the \emph{HST}
WFC3 F140W image (Figure~\ref{fig4}, right).  This pair of bipolar
lobes, hereafter named as the inner lobes, is expected to be 
another hourglass structure located between the two more obvious 
lobes described above.  This smaller eye-shaped feature, which is 
probably due to projection of the inner pair of bipolar lobes, can 
be marginally seen in the H$_{2}$ image, and has been noted in 
the \emph{HST} NICMOS F160W image \citep[][Figure~2 therein]{KH07}.

\begin{figure*}[ht!]
\begin{center}
\includegraphics[width=13.0cm,angle=0]{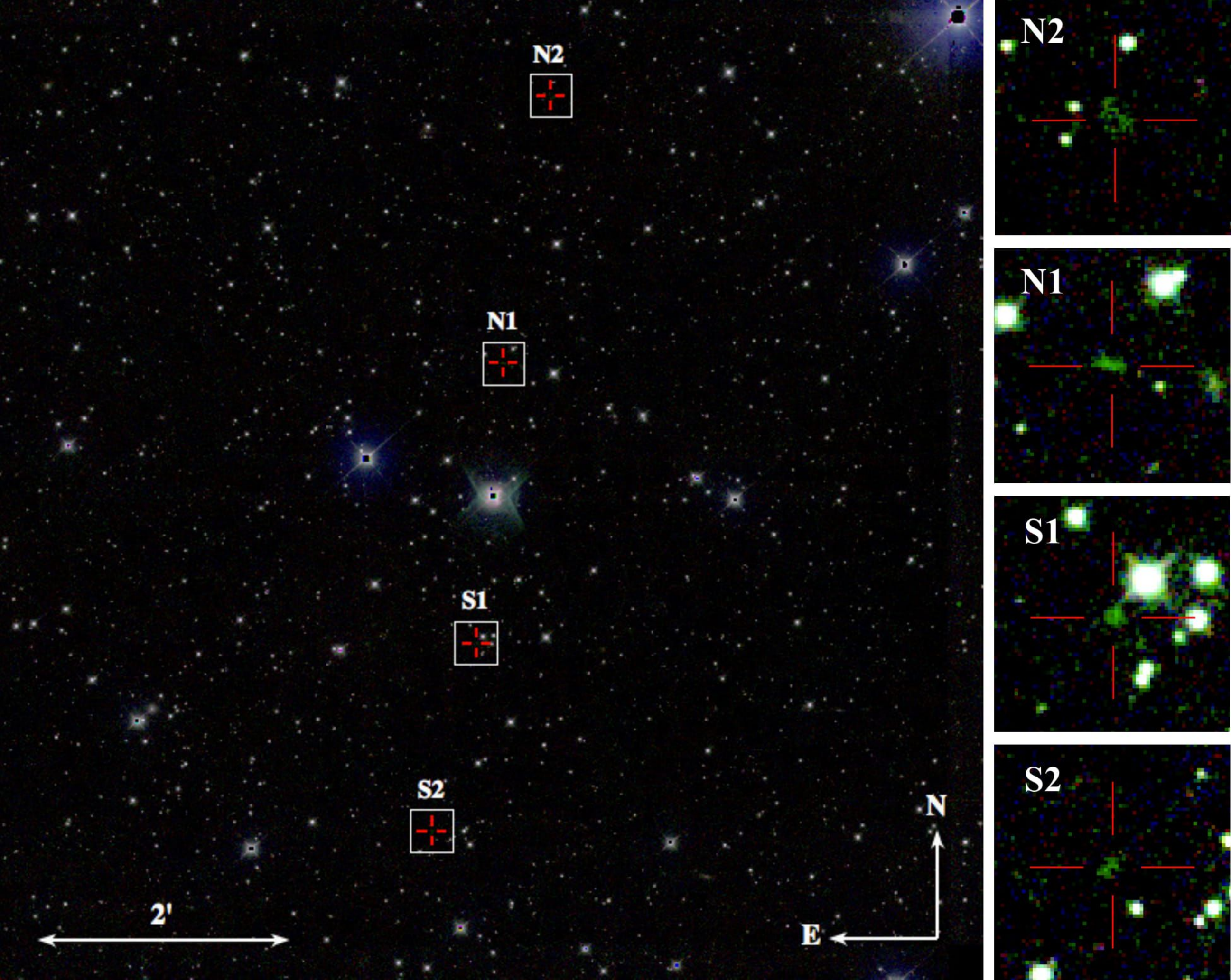}
\caption{Same as Figure~\ref{fig3} but with a wider field centered 
on Hb\,12.  The two pairs of H$_{2}$-emitting knots, defined as S1 
and N1 for the inner pair and S2 and N2 for the outer pair, are 
marked, and zoomed in the right panels, each with a size of 
20\arcsec$\times$20\arcsec.} 
\label{fig5}
\end{center}
\end{figure*}

Moreover, we found in the H$_{2}$ image two pairs of faint, 
point-symmetric knots along the polar axis of Hb\,12 
(Figure~\ref{fig5}, marked as N1 and S1, and N2 and S2) with 
projected lengths of $\sim133$\arcsec\ and $\sim353$\arcsec, 
respectively.  The inner pair of knots (N1 and S1) have been 
detected in [N~{\sc ii}] \citep{Vaytet09}, while the outer knots 
(N2 and S2) are new discoveries.  The two pairs of knots have 
slightly different position angles, probably indicating this PN 
hosts a bipolar, rotating, episodic jet (BRET; C.-H. Hsia et al., 
in preparation).  These features have been briefly reported in 
\citet{Hsia16}.

\subsection{NGC\,6445}
\label{sec3:2}

NGC\,6445 has been classified as a bipolar PN 
\citep[e.g.,][]{Aller73,Peimbert83,Perinotto91,CS95}.  Its He/H and
N/O abundance ratios are consistent with the Type~I definition 
\citep{Peimbert83}.  Previous optical imagery of this PN reveals 
a bright central ring-shaped morphology and open bipolar lobes 
\citep{vanHoof00}; the outer envelope emission of NGC\,6445 is 
[N~{\sc ii}]-dominant \citep{PRL10}.  The physical structure and 
internal kinematics were studied by \citet{Vazquez04}.  Despite 
these multi-wavelength studies, the properties and exact structure 
of NGC\,6445 are still poorly defined \citep{PRL10}.  Our NOT 
ALFOSC wide-field optical images show that NGC\,6445 has an 
irregular-shaped central region with a size of 
$\sim$40\arcsec$\times$50\arcsec, where the [O~{\sc iii}] emission 
dominates, while the [N~{\sc ii}] emission is much more extended 
and defines an overall bipolar morphology (Figure~\ref{fig1}).  The 
bipolar lobes are along the east-west (EW) direction with 
PA$\sim$80\degr, stretching to $\sim$1\farcm6 from center.  There 
are also many filamentary structures across the nebula, which hint 
at a more complex morphology of NGC\,6445 than previously claimed. 
The breadths of the EW lobes reach $\sim$1\farcm8 at both ends. 

\begin{figure*}
\begin{center}
\includegraphics[width=14.0cm,angle=0]{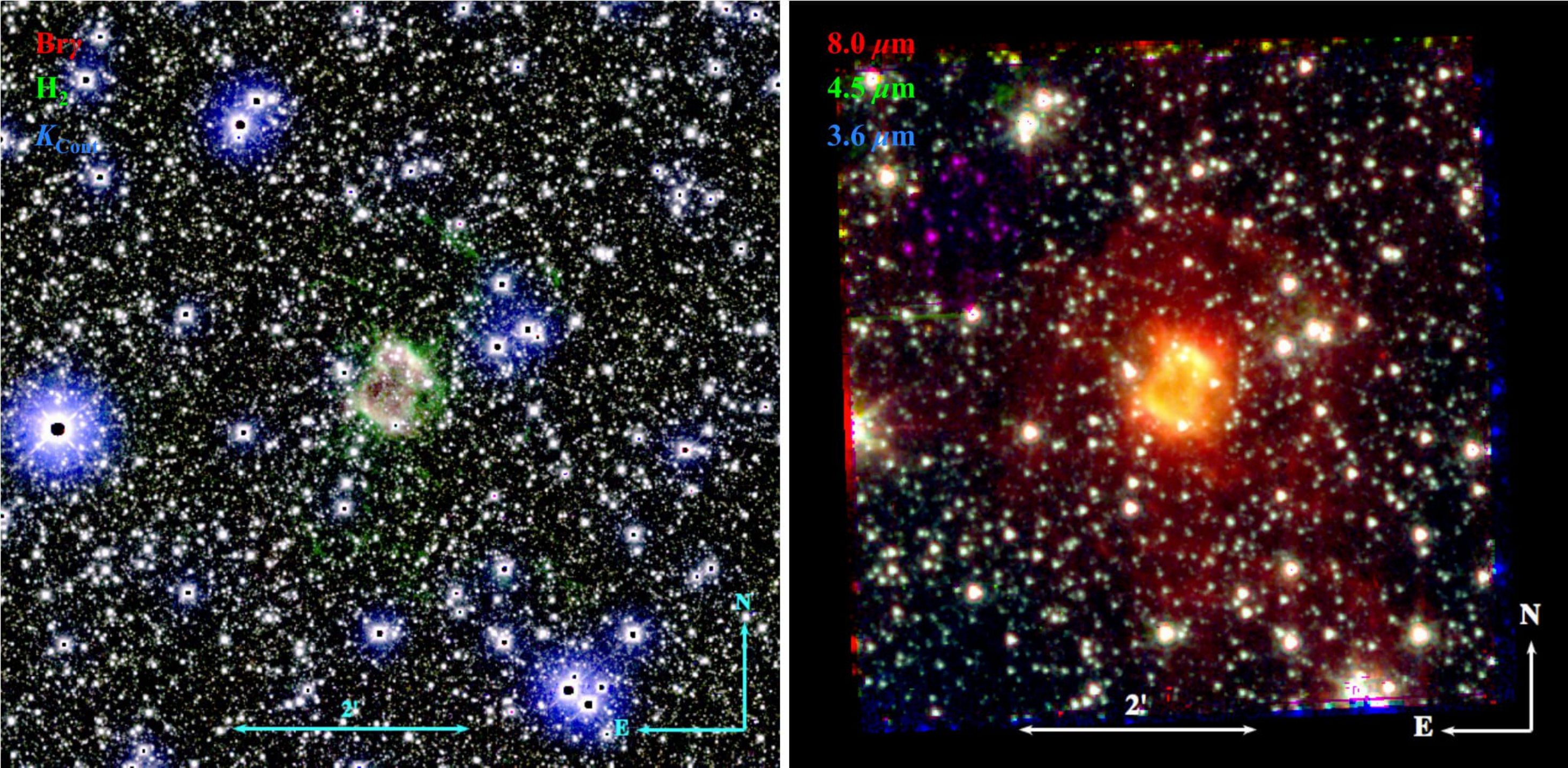}
\caption{Left: CFHT color-composite image of NGC\,6445; color code 
is indicated.  Right: \emph{Spitzer} color-composite image of 
NGC\,6445 created with the IRAC 3.6\,$\mu$m (blue), 4.5\,$\mu$m 
(green) and 8.0\,$\mu$m (red) bands.  Image size and orientation 
are the same as the left panel.} 
\label{fig6}
\end{center}
\end{figure*}

Our CFHT WIRCam images of NGC\,6445 show that the majority of the 
H$_{2}$ 2.122\,$\mu$m, Br$\gamma$ 2.166\,$\mu$m and K$_{\rm c}$ 
2.218\,$\mu$m emission comes from its central region 
(Figure~\ref{fig6}, left).  There are very faint, extended 
structures in the H$_{2}$ emission, which are almost overwhelmed 
by bright field stars but are still marginally visible in the 
color-composite image, where the extended H$_{2}$ emission seems 
to be limb-brightened.  The \emph{Spitzer} mid-IR images show that 
NGC\,6445 is more extended in the 8.0\,$\mu$m emission 
\citep{PRL10}, which fills the halo region defined by the seemingly 
limb-brightened H$_{2}$ emission (Figure~\ref{fig6}, right). 

\begin{figure*}
\begin{center}
\includegraphics[width=14.0cm,angle=0]{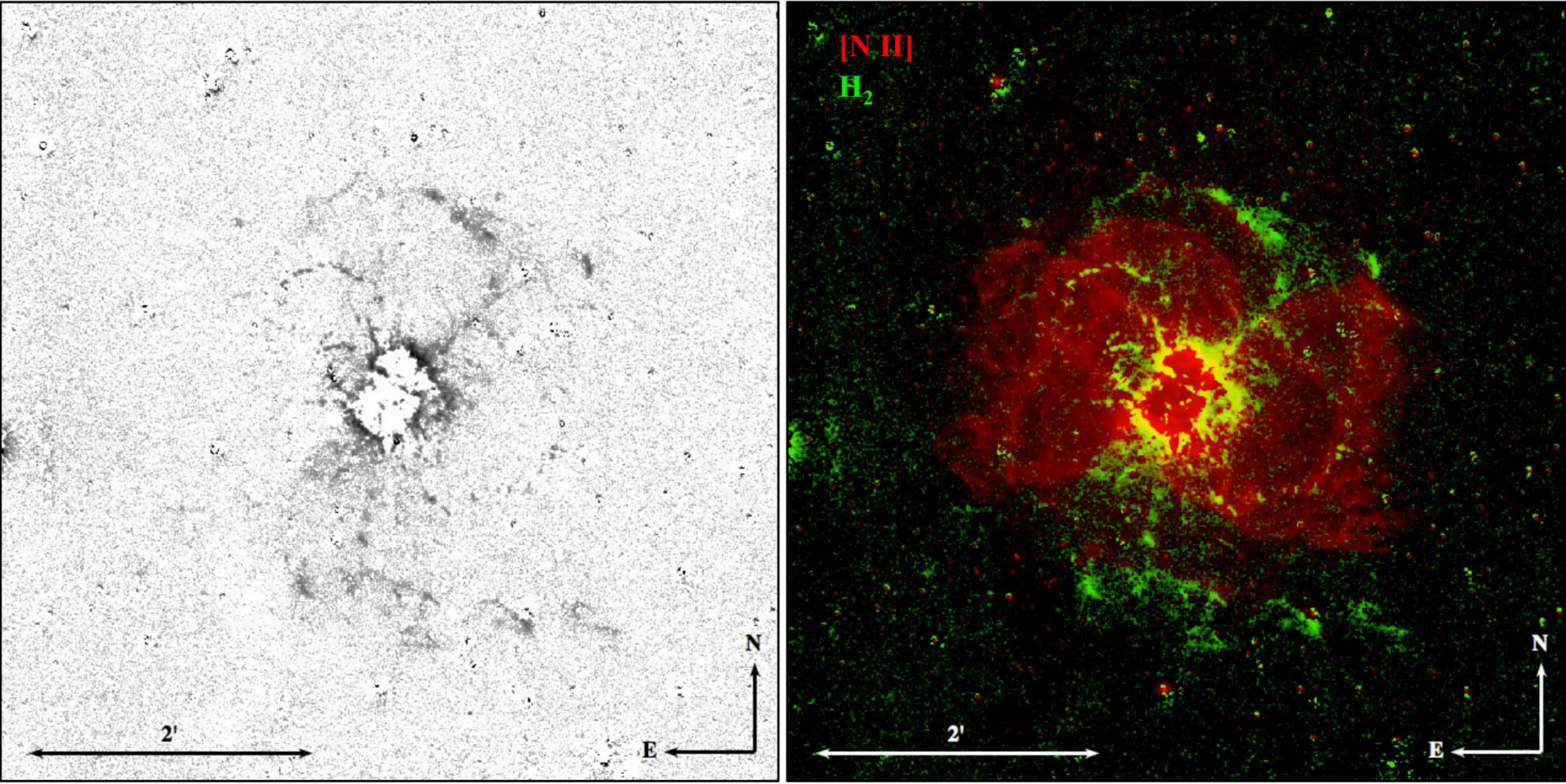}
\caption{Left:  residual H$_{2}$ image of NGC\,6445 created by 
subtracting the scaled Br$\gamma$ image from the H$_{2}$ image. 
Right:  color-composite image of NGC\,6445 created with the 
[N~{\sc ii}] (red) and the residual H$_{2}$ (green) images.} 
\label{fig7}
\end{center}
\end{figure*}

In order to alleviate the effects of the dense stellar field and 
better show the faint halo in H$_{2}$, we created a residual image 
of NGC\,6445 by subtracting the scaled Br$\gamma$ image from 
H$_{2}$.  This residual image is expected to be in H$_{2}$ emission 
only and shows filamentary features well outside the central region 
along the north-south (NS) direction (Figure~\ref{fig7}, left). 
These filaments are located $\sim$1\farcm3--1\farcm6 from the 
centre.  The southern filaments are more extended along the EW 
direction.  There are also rays of faint emission coming from the 
central region. The factor used to scale the Br$\gamma$ image was 
carefully chosen so that emission of stars are mostly removed and 
the nebular structures best seen in the residual image. 

We are aware that (1) it is dangerous to subtract the Br$\gamma$ 
emission from the H$_{2}$ image because H$_{2}$ may be partially 
removed if it is spatially coincident with Br$\gamma$, and (2) it is
conventional to subtract the $K$-continuum emission from the H$_{2}$
image.  However, our main purpose is to search nebular structures 
in H$_{2}$ emission in the outer reaches of previously know PNe; 
besides, the Br$\gamma$ emission is mostly present in the inner 
nebulae, and is much fainter in the outer halos where the H$_{2}$ 
emission is detected.  As mentioned at the beginning of 
Section~\ref{sec3}, the $K_{\rm c}$ filter of WIRCam does not cover 
the real continuum, but include the 1$-$0 S(0) and 2$-$1 S(1) 
emission lines of H$_{2}$.  It is thus even more dangerous to 
subtract the $K_{\rm c}$ image from the H$_{2}$ the, if one wants 
to study the extended structures in the H$_{2}$ emission. 

A close comparison of the residual H$_{2}$ image against the 
\emph{Spitzer} IRAC 8.0\,$\mu$m and the NOT ALFOSC [N~{\sc ii}] 
images shows that along NS, the H$_{2}$ emission generally 
delineates the outer boundary of a broad region where the 
8.0\,$\mu$m emission dominates (Figure~\ref{fig6}, right), while 
the [N~{\sc ii}] emission is more extended along the EW direction,
in consistency with the bipolar morphology, and is confined within
the 8.0\,$\mu$m-emitting region along the NS.  Along the NS 
direction, the H$_{2}$ emission appears to be slightly further out 
than the [N~{\sc ii}] emission (Figure~\ref{fig7}).  This reinforces
the notion that the optical emission may be significantly affected 
by the illumination of UV photons and cannot represent the intrinsic
matter distribution of PNe \citep[e.g.,][]{Zhang12c}.  The [N~{\sc 
ii}] image mainly traces the bipolar lobes, while the H$_{2}$ image
reveals the limb-brightened gas extended along the equatorial 
direction. 


The northern and the southern H$_{2}$-emitting filaments, 
$\sim$1\farcm4--1\farcm6 from the centre of NGC\,6445, might be the 
outer boundaries of an edge-on-viewed torus of this PN.  This 
torus was ejected in the AGB phase and is now being disrupted due 
to interaction with the fast stellar wind that was developed later 
on.  The south region of the torus seems to be much more disrupted 
than its north counterpart.  Within the EW bipolar lobes, there 
seem to be two arcs in the H$_{2}$ emission, as shown in the 
residual image (Figure~\ref{fig7}): a north-east (NE) arc and a 
south-west (SW) one, both $\sim$1\farcm0 from the nebular centre. 
The SW arc seems to be disrupted, but matches the position of a 
giant arc in [N~{\sc ii}].  These two H$_{2}$ arcs might define 
another pair of bipolar lobes, which have a PA$\sim$56\degr.  This 
PA generally agrees with that of the [O~{\sc iii}]-right inner 
nebular region of NGC\,6445 (Figure~\ref{fig1}), suggesting that 
the smaller bipolar bubbles are of relatively higher excitation, 
and thus might have developed recently.

\subsection{NGC\,6543}
\label{sec3:3}

Morphological and structural studies of NGC\,6543 (a.k.a., the 
Cat's Eye Nebula) based on the optical images have been carried out 
extensively \citep[e.g.,][]{Balick92,HB94,Reed99,Corradi03,Balick04,
BH04}.  A giant (angular diameter $\sim$300\arcsec) 
patchy/filamentary outer halo has previously been discovered beyond 
the bright core in the optical wavebands \citep{Mil74,Middle89,
Balick92}.  The \emph{HST} imagery revealed regularly spaced, 
concentric circular rings surrounding the bright core 
(Figure~\ref{fig1}), which were explained as spherical bubbles of 
periodic isotropic mass pulsations that preceded the formation of 
the bright nebular core \citep{Balick01}.

\begin{figure*}
\begin{center}
\includegraphics[width=14.0cm,angle=0]{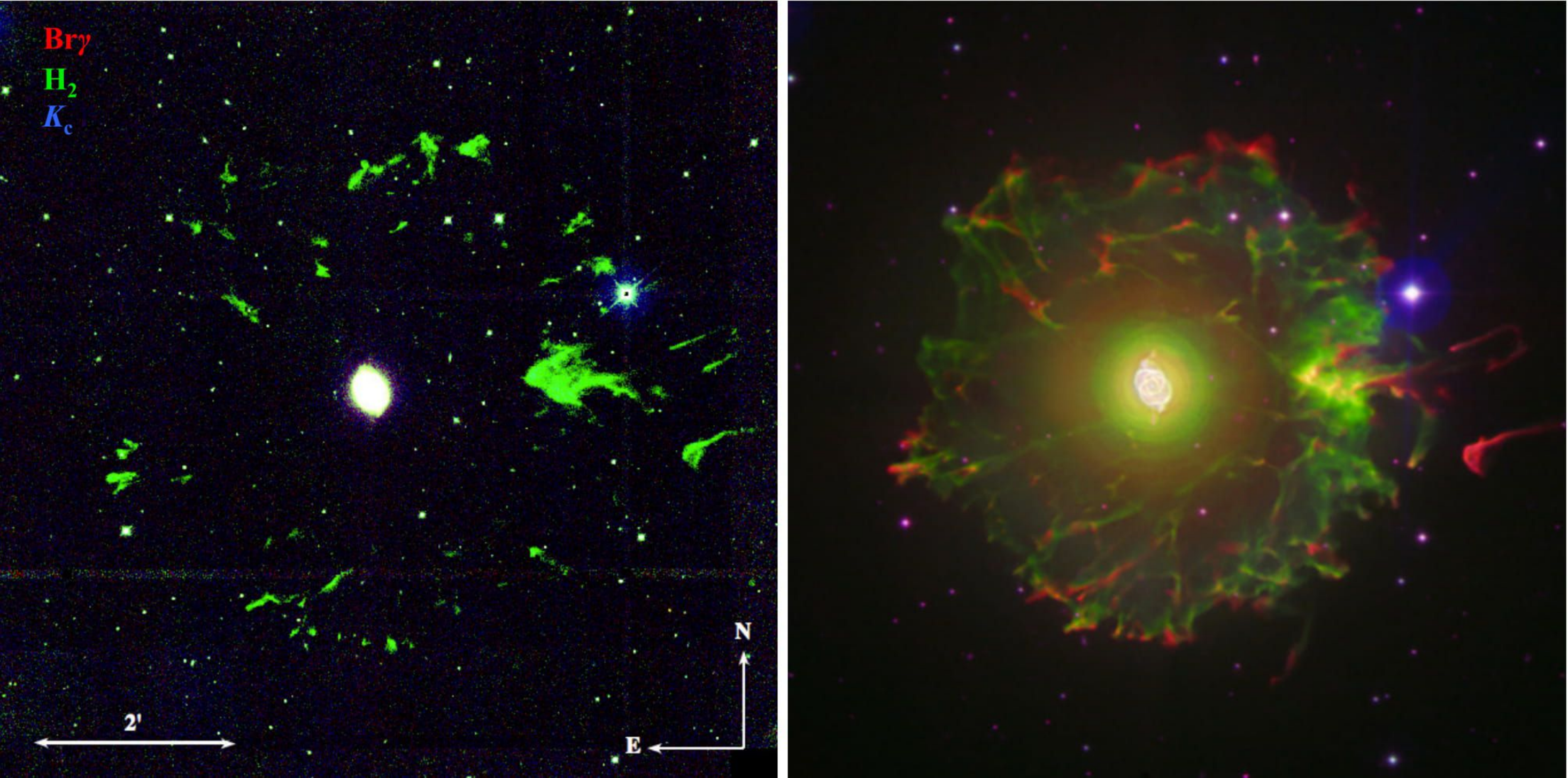}
\caption{Left: CFHT color-composite image of NGC\,6543.  Right: 
optical image of NGC\,6543 obtained from the 2.5\,m Isaac Newton 
Telescope; adopted from the Isaac Newton Group of Telescopes 
(http://www.ing.iac.es/PR/press; image credit: D.\ L\'{o}pez); 
red is the [S~{\sc ii}] line emission, green is H$\alpha$, and 
blue is [O~{\sc iii}].  Image size and orientation are the same 
as left.} 
\label{fig8}
\end{center}
\end{figure*}

\begin{figure}
\begin{center}
\includegraphics[width=7.0cm,angle=0]{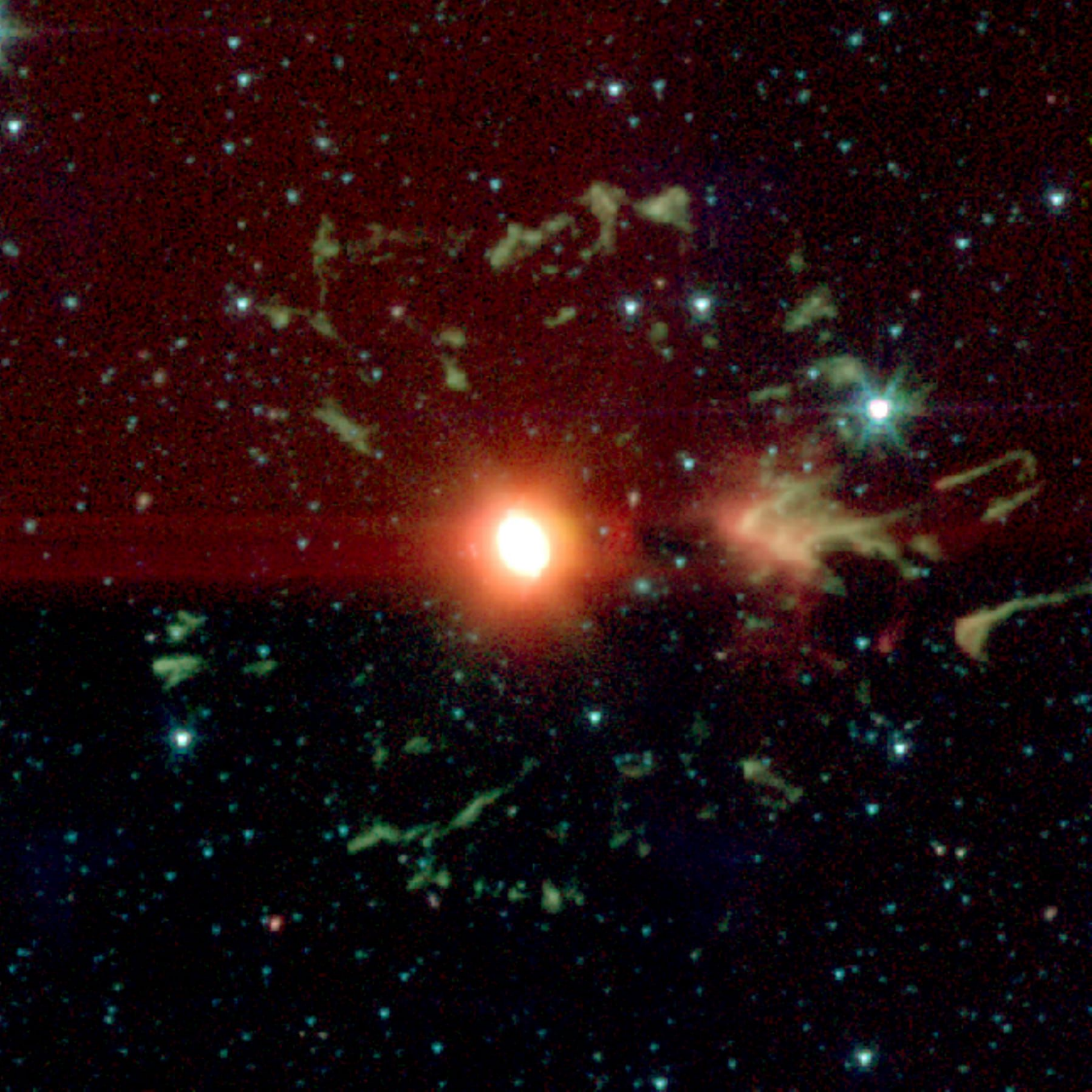}
\caption{\emph{Spitzer} color-composite image of NGC\,6543 created 
with the IRAC 3.6\,$\mu$m (blue), 4.5\,$\mu$m (green) and 
8.0\,$\mu$m (red) bands.  Image size and orientation are the same 
as Figure~\ref{fig8}.} 
\label{fig9}
\end{center}
\end{figure}

Our CFHT near-IR images show that the giant halo surrounding the 
Cat's Eye is dominated by H$_{2}$ emission (Figures~\ref{fig8}, 
left), which generally follows [S~{\sc ii}] that delineates the 
outer edges of the knotty/filamentary features (Figure~\ref{fig8}, 
right); the morphologies of the H$_{2}$ features also follow the 
fine structures observed in H$\alpha$+[N~{\sc ii}] 
\citep[][Figure~4 therein]{Balick92}, except that the halo seems 
smoother in the latter broad-band filter.  This morphology 
indicates that H$_{2}$ (and also [S~{\sc ii}]) may be shock 
excited.  Previous optical spectroscopy found that the [O~{\sc 
iii}] electron temperature of the faint halo of NGC\,6543 is higher
than that of the bright central core; this was proposed to be due 
to hot fast stellar wind that escaped from the central nebula and 
shocked the filamentary halo \citep{Middle89,Middle91,Meaburn92}. 
This fast wind mass-loads as it penetrates through the central 
nebula, resulting in a mildly supersonic flow that shocks the 
clumps of material in the halo region \citep{Dyson92}.  The halo 
is also observable in \emph{Spitzer} (Figure~\ref{fig9}).  The 
spatial distribution of dust emission resembles that of H$_2$.  It 
is still unclear whether there is relevance between the halo and 
the concentric circular rings around the inner Cat's Eye 
(Figure~\ref{fig1}). 

The filamentary outer halo of NGC\,6543 were observed to be detached
from the concentric rings around the bright inner Cat's Eye, 
indicating that the halo material was probably ejected well before 
the periodic ejection of the spherical gas.  The average kinematical
age of this giant halo, as computed assuming a constant expansion 
velocity, is $\sim$85\,000~yr \citep{Corradi03}.  Based on the 
spacing between the concentric rings and an assumption of constant 
expansion velocity (10~km~s$^{-1}$), \citet{Balick01} estimated 
that launching of the spherical bubbles from the surface of the 
progenitor star began about 15\,000~yr ago.  For comparison, the 
expansion age of the inner nebula is $\sim$1000~yr \citep{Reed99}. 
Such huge difference in the kinematical ages of the giant halo and 
the circular rings indicates that these two structural components 
could be ejected at different episodes during the evolution from 
the late-AGB to the PN phase.

\subsection{NGC\,6720}
\label{sec3:4}

The halo around NGC\,6720 (a.k.a., the Ring Nebula) has been 
observed in the optical emission 
\citep[e.g.,][]{Balick92,Corradi03} and IR emission lines of 
molecular hydrogen \citep[e.g.,][]{BPG78} and CO 
\citep[e.g.,][]{HH86}.  The central star of NGC\,6720 has exhausted
the hydrogen shell burning and is now on the cooling track.  Thus 
the outer halo of NGC\,6720 is recombining.  The \emph{Herschel} 
far-IR observations reveals a resemblance between the dust 
distribution and the H$_{2}$ emission in the inner nebula, 
suggesting that H$_{2}$ forms on grain surfaces \citep{vanHoof10}. 

\begin{figure*}
\begin{center}
\includegraphics[width=14.0cm,angle=0]{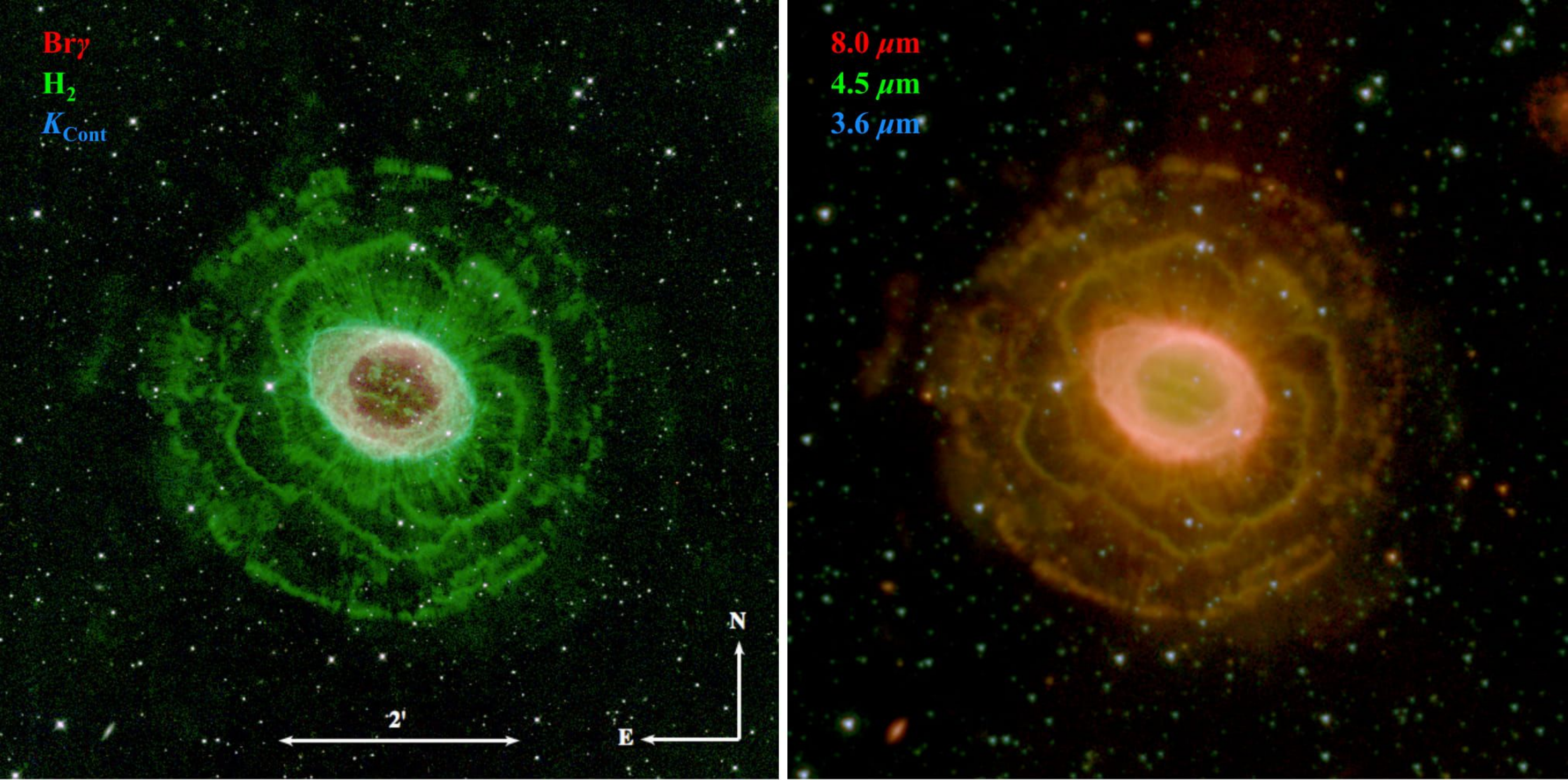}
\caption{Left: CFHT color-composite image of NGC\,6720.  Right: 
\emph{Spitzer} IRAC color-composite image; image size and 
orientation are the same as left.}
\label{fig10}
\end{center}
\end{figure*}

There are structures in the halo of NGC\,6720: the two inner layers 
have petal-like morphologies, while the outermost one is almost 
circular.  These features, together with the radial filaments 
emerging from the central nebula, point to a possibility that the 
recombining halo of NGC\,6720 might be in the process of disruption 
due to interaction with the stellar wind.  The halo is dominated in 
the H$_{2}$ emission and also visible in $K_{\rm c}$; this may imply 
a relatively low H$_{2}$ 1$-$0 S(1) to H$_{2}$ 2$-$1 S(1) line 
ratio, but it is difficult to resolve quantitatively because of many 
contributions (from stellar continuum, dust continuum, and H$_{2}$ 
lines) to this $K_{\rm c}$ filter.  If the H$_{2}$ 
1$-$0\,S(1)/2$-$1\,S(1) ratio is indeed low in the halo of 
NGC\,6720, its H$_{2}$ emission is then likely due to UV pumping 
\citep[e.g.,][]{Marquez15}, especially for the petal-like features 
of the inner halo that disclose regions where UV photons escape from 
the core region.  The \emph{Spitzer} IRAC near-IR images show that 
the $\sim$60\arcsec$\times$88\arcsec\ elliptical-shaped inner region 
of NGC\,6720 is very bright in the 8.0\,$\mu$m emission.  The 
8.0\,$\mu$m emission in the outer halos is also stronger than the 
other three IRAC bands (Figure~\ref{fig10}, right). 

\begin{figure}
\begin{center}
\includegraphics[width=7.0cm,angle=0]{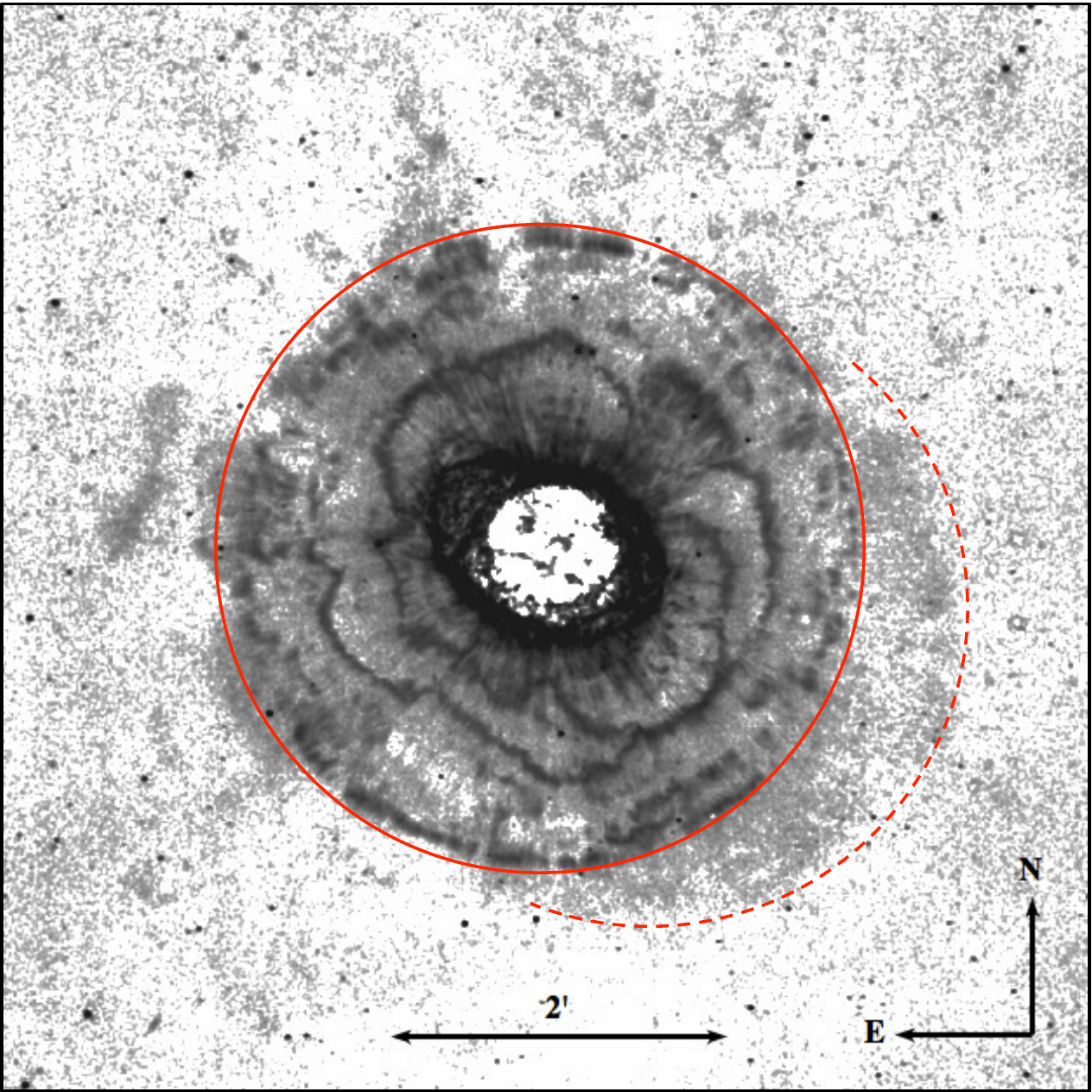}
\caption{Residual image of NGC\,6720 created by subtracting the
scaled Br$\gamma$ image from the H$_{2}$ image.  The central 
region is saturated to enhance the extended structure.  The red 
solid circle is a schematic fit to the $\sim$2\arcmin-radius outer 
halo, and the red dashed arc delineates the faint southwest 
crescent.} 
\label{fig11}
\end{center}
\end{figure}

We created a residual image of NGC\,6720 using the same method as 
we did for NGC\,6445 (Section~\ref{sec3:2}) to enhance the extended 
morphologies in H$_{2}$.  The radial filaments, pointing away from 
the central ring and penetrating through the inner (petal-like) 
halo, is clearly visible in Figure~\ref{fig11}.  Besides, a 
crescent-shaped structure was found attached to the southwest of 
the circular outer halo; this might be due to inclination of the 
bipolar lobes of NGC\,6720, as suggested by \citep{Kast96}. 
\citet{Kwok08} speculate a triple biconic structure of the outer 
halos of NGC\,6720.

An alternative model was proposed by \citet[][Figure~14 
therein]{Guerrero97} who, through high-dispersion spectroscopy, 
suggest that the bright ring of NGC\,6720 is actually a closed 
shell with a prolate ellipsoidal geometry with density enhancement
near the equator, and surrounding this inner shell there is a halo 
of remnant red giant wind.  Based on the \emph{HST} imaging, 
\citet[][Figure~11 therein]{ODell13} suggest that the main ring of 
NGC\,6720 is an ionization-bounded disk with a central cavity and 
perpendicular extended lobes pointed nearly along the line of 
sight.  Detailed morpho-kinematic analysis with the aid of models 
is needed to assess these three interpretations of NGC\,6720's 
intrinsic structure.

\subsection{NGC\,6772}
\label{sec3:5}

\begin{figure*}
\begin{center}
\includegraphics[width=14.0cm,angle=0]{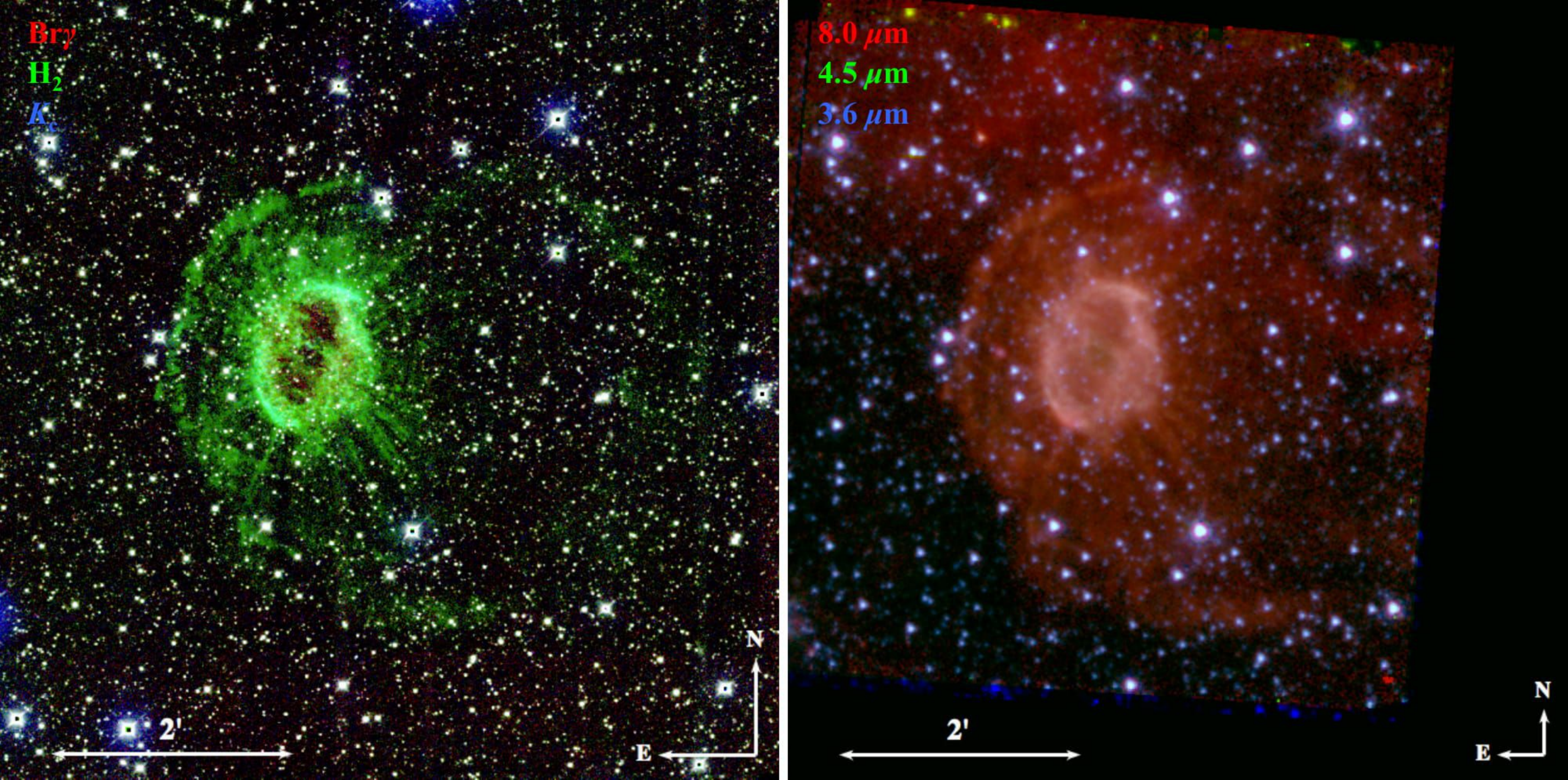}
\caption{Left: CFHT color-composite image of NGC\,6772.  Right: 
\emph{Spitzer} IRAC color-composite image; image size and 
orientation are the same as left.} 
\label{fig12}
\end{center}
\end{figure*}

NGC\,6772 has an obviously asymmetric halo in H$_{2}$ emission,
but invisible in Br$\gamma$ (Figure~\ref{fig2}).  This halo is 
better shown in Figure~\ref{fig12} (left), where we can see that 
H$_{2}$ emission dominates the halo, while Br$\gamma$ only comes 
from the inner region.  The H$_{2}$ emission is mainly concentrated 
on the inner nebular shell, which has a size of 
$\sim$28\arcsec$\times$37\arcsec\ and is slightly tilted (with 
PA$\sim$165$\degr$).  The distribution of Br$\gamma$ emission is 
more homogeneous within the nebular shell, and is well consistent 
with H$\alpha$ (Figure~\ref{fig1}).  It is interesting to note that 
H$_{2}$ generally traces [N~{\sc ii}]:  they both mainly come from 
the inner shell.  Within this shell, there are also some H$_{2}$ 
filaments that seem to be also present in the [N~{\sc ii}] image. 

Although previously studied in the mid- and near-IR \citep[including
the H$_{2}$ 2.122\,$\mu$m line; e.g.,][]{Webster88,Kast94,RLP09,
Marquez13}, the complete morphology of NGC\,6772's halo is so far 
only displayed in our CFHT images.  The \emph{Spitzer} IRAC imaging 
reported in the literature does not cover the whole SW halo of 
NGC\,6772 (Figure~\ref{fig12}, right).  The \emph{Spitzer} 
8.0\,$\mu$m emission generally traces the extended halo of 
NGC\,6772, but its spatial resolution is much lower than our CFHT 
WIRCam images. 

\begin{figure*}
\begin{center}
\includegraphics[width=14.0cm,angle=0]{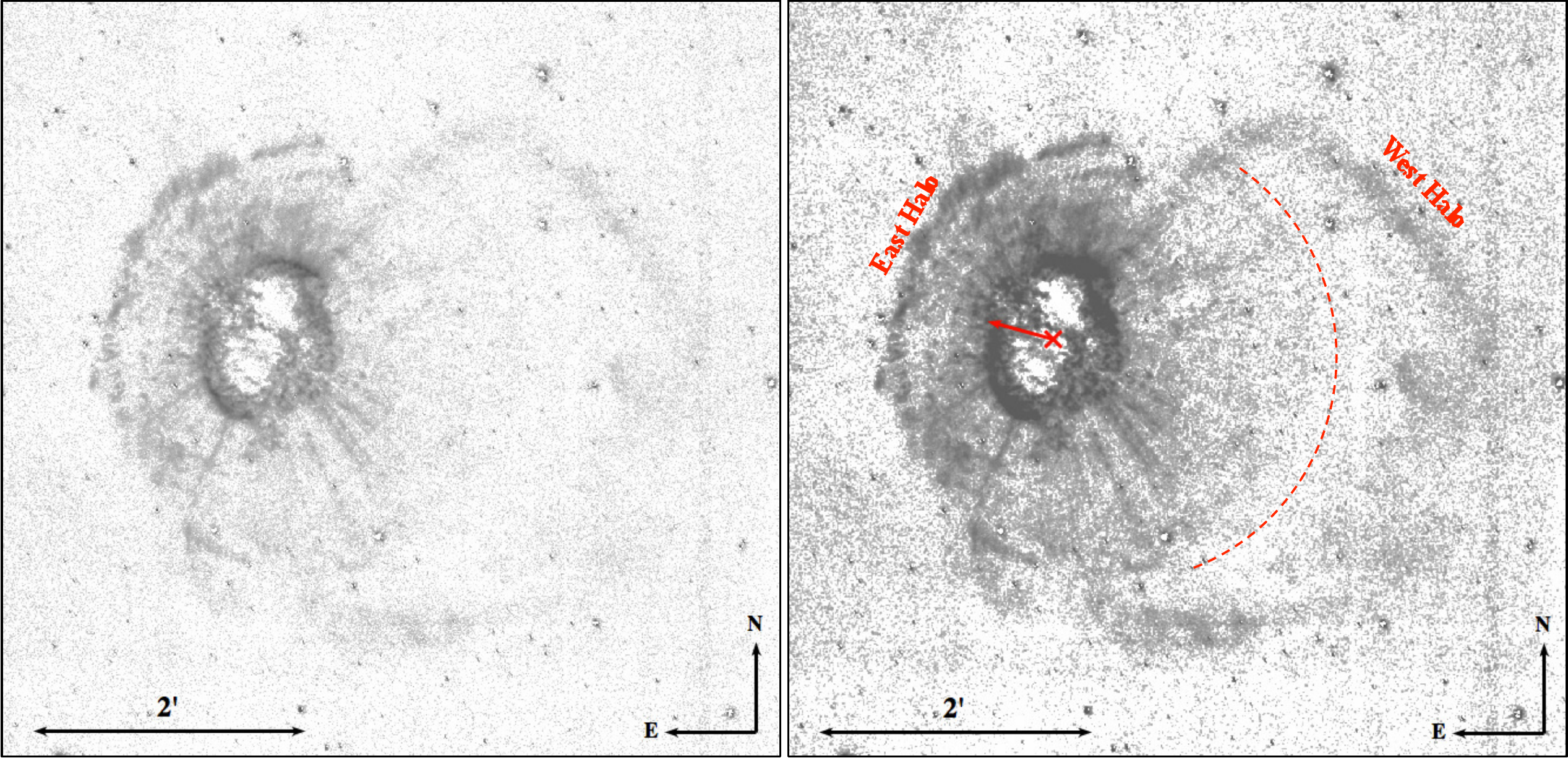}
\caption{Left: residual image of NGC\,6772 created by subtracting 
the scaled Br$\gamma$ image from the H$_{2}$ image. 
Right: same as left but with a different cut level and slightly 
Gaussian smoothed to enhance the extended halo structure. The East 
and the West Halos may originally belong to the same circular outer
halo.  The red dashed arc roughly delineates the boundary of the 
disrupted inner halo.  The red cross marks the central star 
position, and the red arrow with PA=77$\degr$ indicates the 
possible direction of proper motion.} 
\label{fig13}
\end{center}
\end{figure*}

In the residual H$_{2}$ image of NGC\,6772 (Figure~\ref{fig13}), 
we found that the eastern halo, although slightly fragmented, is 
overall bow-shaped and its outer boundary is 
$\sim$75\arcsec--80\arcsec\ from the central star, while the 
western halo seems to have been disrupted and extends as far as 
$\sim$3\farcm2 from the central star.  This morphology indicates 
that NGC\,6772 might be interacting with the ISM.  NGC\,6772 thus 
best shows the possible PN-ISM interaction among our sample. 
According to the morphology of the extended halo, we conclude that 
this PN may have a proper motion with respect to the local ISM at 
PA$\sim$77$\degr$ (Figure~\ref{fig13}, right). 
Tn the H$_{2}$ image of NGC\,6772 is that there are numerous radial 
filaments/rays coming out from the inner shell (or, the inner 
ring), pointing toward the outer halo; they might be shadows from 
the clumps in the inner shell.  Such radial features are also seen 
in several other PNe, most obviously in NGC\,6720 
(Figures~\ref{fig2}, \ref{fig10} and \ref{fig11}).

\subsection{NGC\,6781}
\label{sec3:6}

Optical images of NGC\,6781 obtained in [O~{\sc iii}], H$\alpha$ 
and [N~{\sc ii}] (Figure~\ref{fig1}) show a slightly elliptical, 
[O~{\sc iii}]-luminous inner region with a radius of 
$\sim$50\arcsec, and an extended nebular structure along the 
north-south direction, where [N~{\sc ii}] emission is dominant. 
The inner nebular shell is also enhanced in [N~{\sc ii}] emission. 
This morphology was observed by \citet{Phil11}, who studied this 
PN using the NOT$+$\emph{HST} images as well as the \emph{Spitzer} 
mid-IR and \emph{ISO} near-IR data. 

\begin{figure*}
\begin{center}
\includegraphics[width=14.0cm,angle=0]{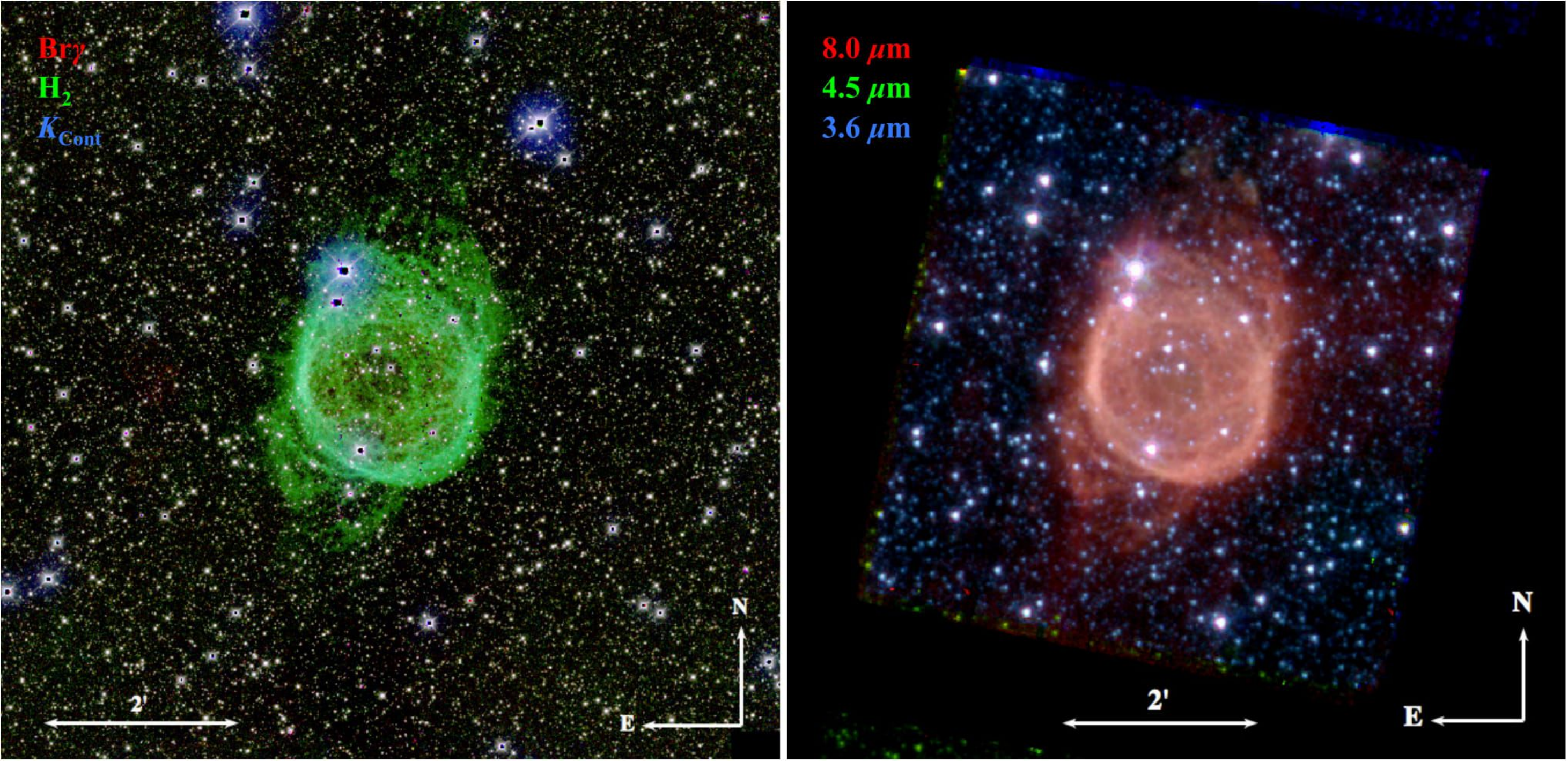}
\caption{Left: CFHT color-composite image of NGC\,6781.  Right: 
\emph{Spitzer} IRAC color-composite image; image size and 
orientation are the same as left.} 
\label{fig14}
\end{center}
\end{figure*}

Our deep H$_{2}$ 2.122\,$\mu$m image shows that NGC\,6781 has a 
slightly elliptical-shaped central nebula and extended nebular 
structures along PA$\sim$120$\degr$ (Figure~\ref{fig14}). There are 
a series of arcs along the south extended region, with their 
intensities decreasing further out. The northern extended structure 
is more diffuse than its southern counterpart.  This morphology is 
generally consistent with the [N~{\sc ii}] image.  More details are 
clearly revealed for the inner region by our high-resolution image: 
there are many filamentary fine structures in the inner nebula, and 
the H$_{2}$ emission is enhanced at the outer boundary.  The much 
weaker Br$\gamma$ emission mainly comes from the central region and 
has a more homogeneous distribution; the $K_{c}$ emission, although 
weaker, shows similar morphology as seen in H$_{2}$.  In general, 
H$_{2}$ is the dominant emission among the CFHT near-IR bands 
utilized in our observations, and delineates the main structure of 
NGC\,6781. 

\begin{figure*}
\begin{center}
\includegraphics[width=14.0cm,angle=0]{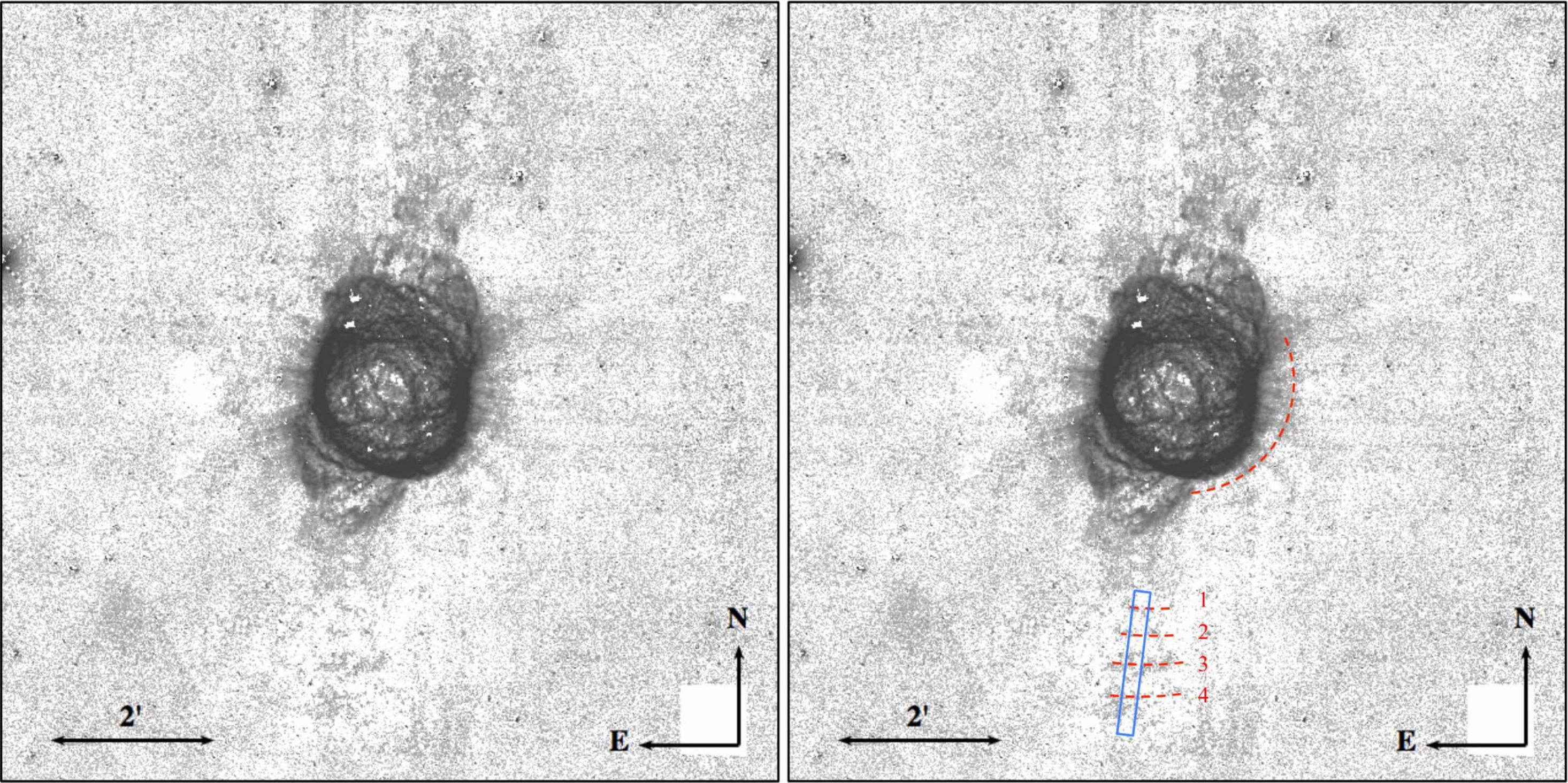}
\caption{Left: residual image of NGC\,6781 created by subtracting 
the scaled Br$\gamma$ image from the H$_{2}$ image and then 
slightly Gaussian smoothed to reduce noise. The central nebula/ring 
is slightly saturated to enhance the extended structure.  Right: 
same as the left panel but over-plotted with schematic sketch of 
possible rings (red dashed curves; see the text for description). 
The blue rectangle is a cut through the four possible rings as 
labeled with numbers.} 
\label{fig15}
\end{center}
\end{figure*}

\begin{figure}
\begin{center}
\includegraphics[width=1.0\columnwidth,angle=0]{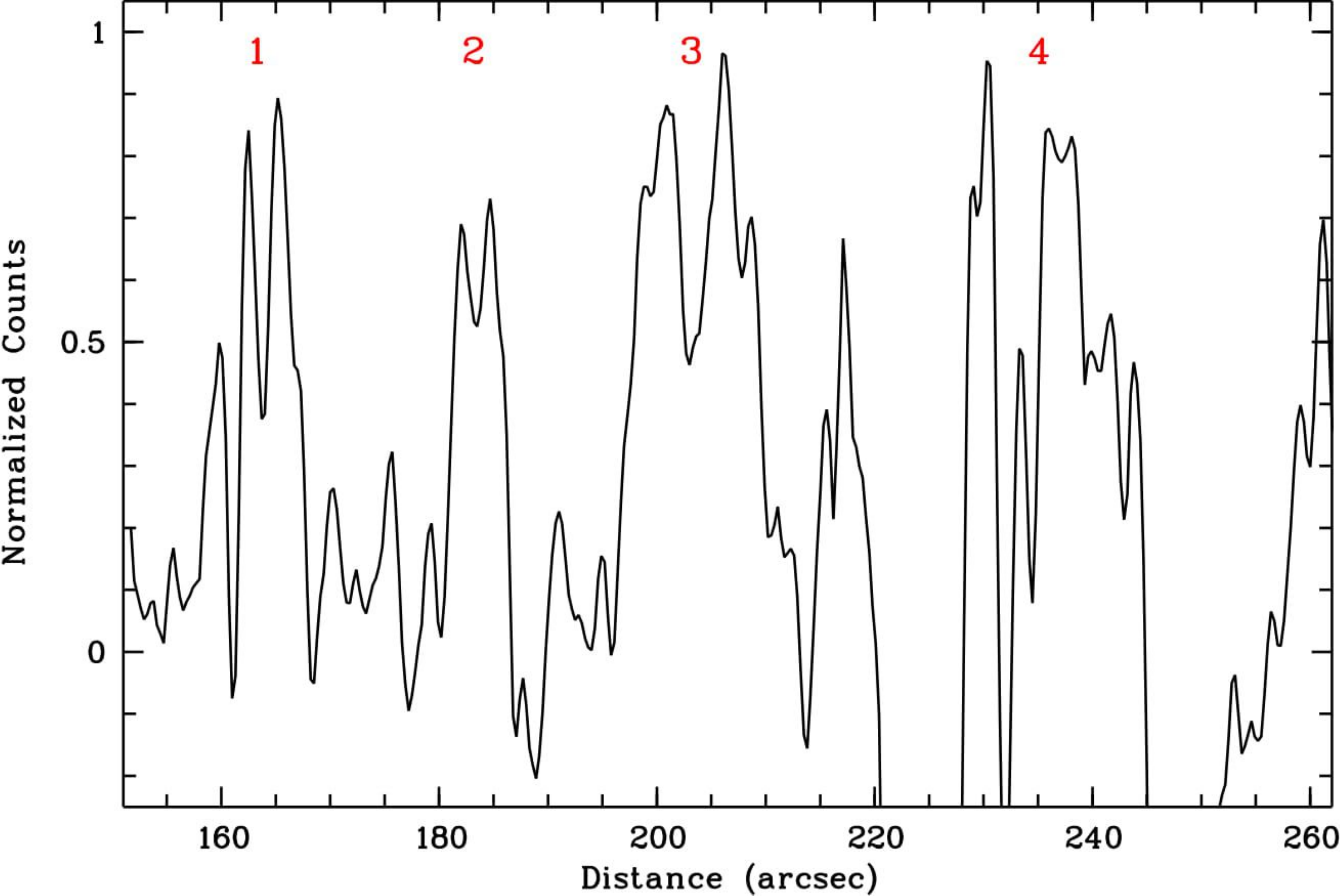}
\caption{Radial profile of the cut through the southern halo of 
NGC\,6781 (the blue rectangle in Figure~\ref{fig15}).  Approximate 
peak positions of the four possible southern rings (labeled with 
numbers in Figure~\ref{fig15}, right) are marked with corresponding 
numbers.  Horizontal axis is the distance (in arcsec) from the 
central star.} 
\label{fig16}
\end{center}
\end{figure}

In order to remove background stars and search for possible extended
structures of NGC\,6781, we created a residual image of NGC\,6781 
(Figure~\ref{fig15}).  In this ``cleaned'' residual H$_{2}$ image, 
we see many radial filaments coming out from the central 
nebula/ring, features also seen in NGC\,6720 and NGC\,6772 
(Figures~\ref{fig10}--\ref{fig13}).  Very close to the inner nebula,
there seems to be a faint ring in the SW (Figure~\ref{fig15}), 
which is $\sim$1\farcm37 from the centre of the PN.  In the south 
halo, we found very faint ``ripples'' that are $\sim$2\farcm7, 
3\farcm0, 3\farcm35, and 3\farcm75 from the centre of the nebula 
(Figure~\ref{fig15}, right).  These faint features might be part 
of the concentric rings in the halo of NGC\,6781.  Although the 
S/N is low, a radial cut through this region does show four peaks 
(Figure~\ref{fig16}).  We did not see any northern counterparts of 
these faint ripples.

\begin{figure}
\begin{center}
\includegraphics[width=7.0cm,angle=0]{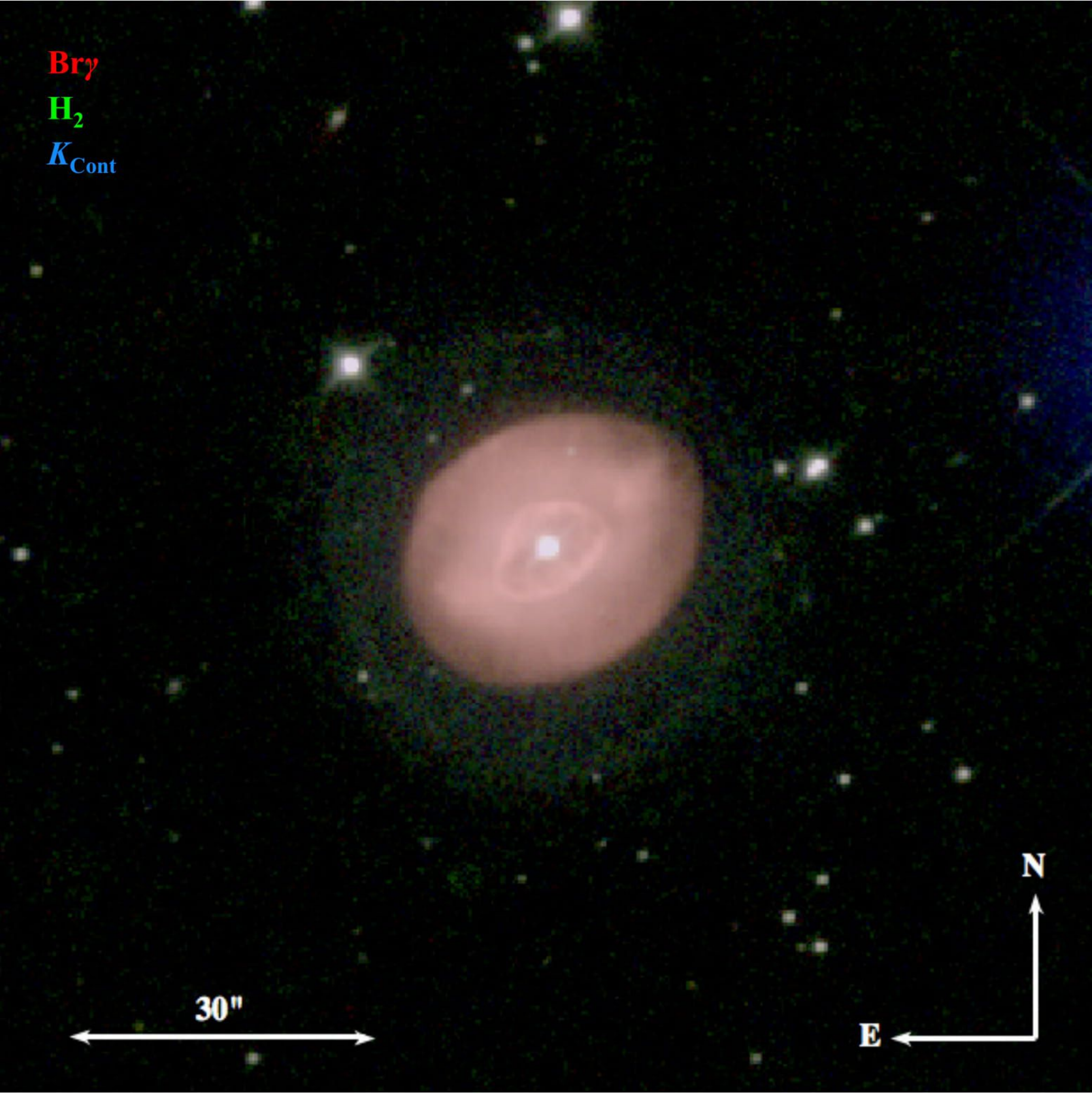}
\caption{CFHT color-composite image of NGC\,6826.} 
\label{fig17}
\end{center}
\end{figure}

\begin{figure}
\begin{center}
\includegraphics[width=7.0cm,angle=0]{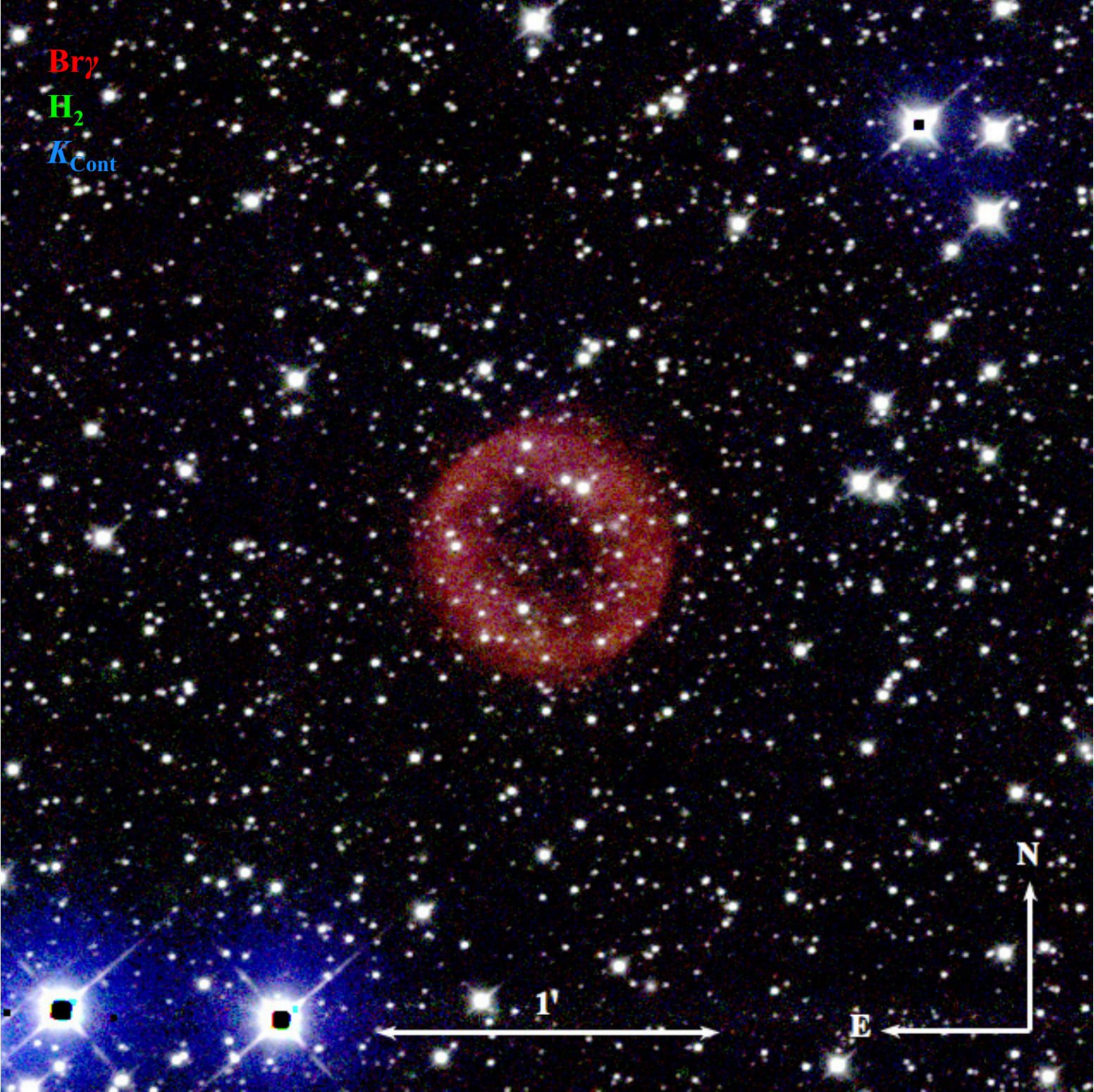}
\caption{CFHT color-composite image of NGC\,6894.} 
\label{fig18}
\end{center}
\end{figure}

The morphology of the inner nebular ring of NGC\,6781, as well as 
the SE H$_{2}$ arcs, has been explained as projected cylindrical 
cavities \citep{Kast94}.  In order to explain the kinematic 
structure of NGC\,6781 observed through high-resolution spectroscopy
of the H$_{2}$ 2.122\,$\mu$m emission line, \citet{Hiriart05} also 
suggested a thin hollow cylindrical shell whose axis is tilted with 
respect to the line of sight.  The faint features discovered in the 
extended halo may help to constrain the 3D structure of NGC\,6781, 
and shed light on the history of the fast stellar wind and the 
material expelled in the AGB phase.

\subsection{NGC\,6826}
\label{sec3:7}

An extended halo, nearly twice the radius of the elliptical-shaped 
inner nebula, has been noticed in the optical images of NGC\,6826 
\citep{Balick92,Corradi03}.  A faint, round halo can also be seen 
in all CFHT filters; it has a sharp boundary extending to 
$\sim$25\arcsec\ from the central star (Figure~\ref{fig17}).  The 
elliptical shaped inner nebula, as defined in the optical bands 
(Figure~\ref{fig1}), is bright in the Br$\gamma$ emission.

\subsection{NGC\,6894}
\label{sec3:8}

No extended nebular structures were found in the halo of NGC\,6894 
in our CFHT images (Figure~\ref{fig2}).  The main nebula shell, 
which has an outer radius of $\sim$24\arcsec, is dominated by the 
Br$\gamma$ emission (Figure~\ref{fig18}).  There is a non-spherical 
central hollow, which is elongated along PA$\sim$35$\degr$.  The 
\emph{HST} broad-band image (Figure~\ref{fig1}) shows that there 
seems to be a ring of halo surrounding the main nebula and extending
to the same size as the CFHT WIRCam images.  It is worth mentioning 
that some detached tail-like ISM material close to NGC\,6894 was 
found \citep{Chu87}. It was suggested that the halo of NGC\,6894 
might have been stripped by the ISM \citep{SZ97}.

\subsection{NGC\,7009}
\label{sec3:9}

\begin{figure}
\begin{center}
\includegraphics[width=7.0cm,angle=0]{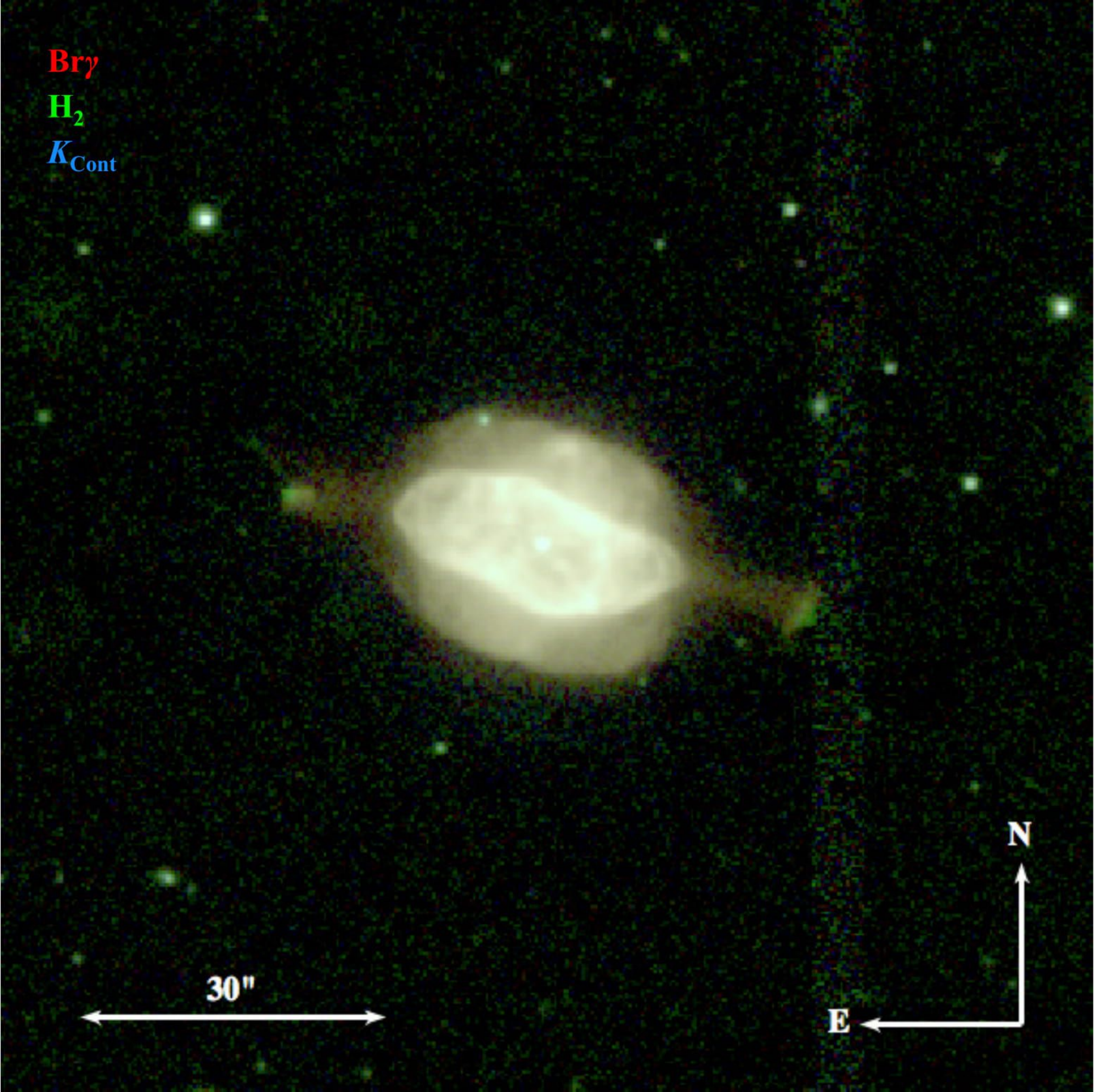}
\caption{CFHT color-composite image of NGC\,7009. The two E-W knots
are enhanced in H$_{2}$ emission.} 
\label{fig19}
\end{center}
\end{figure}

\begin{figure*}
\begin{center}
\includegraphics[width=13.0cm,angle=0]{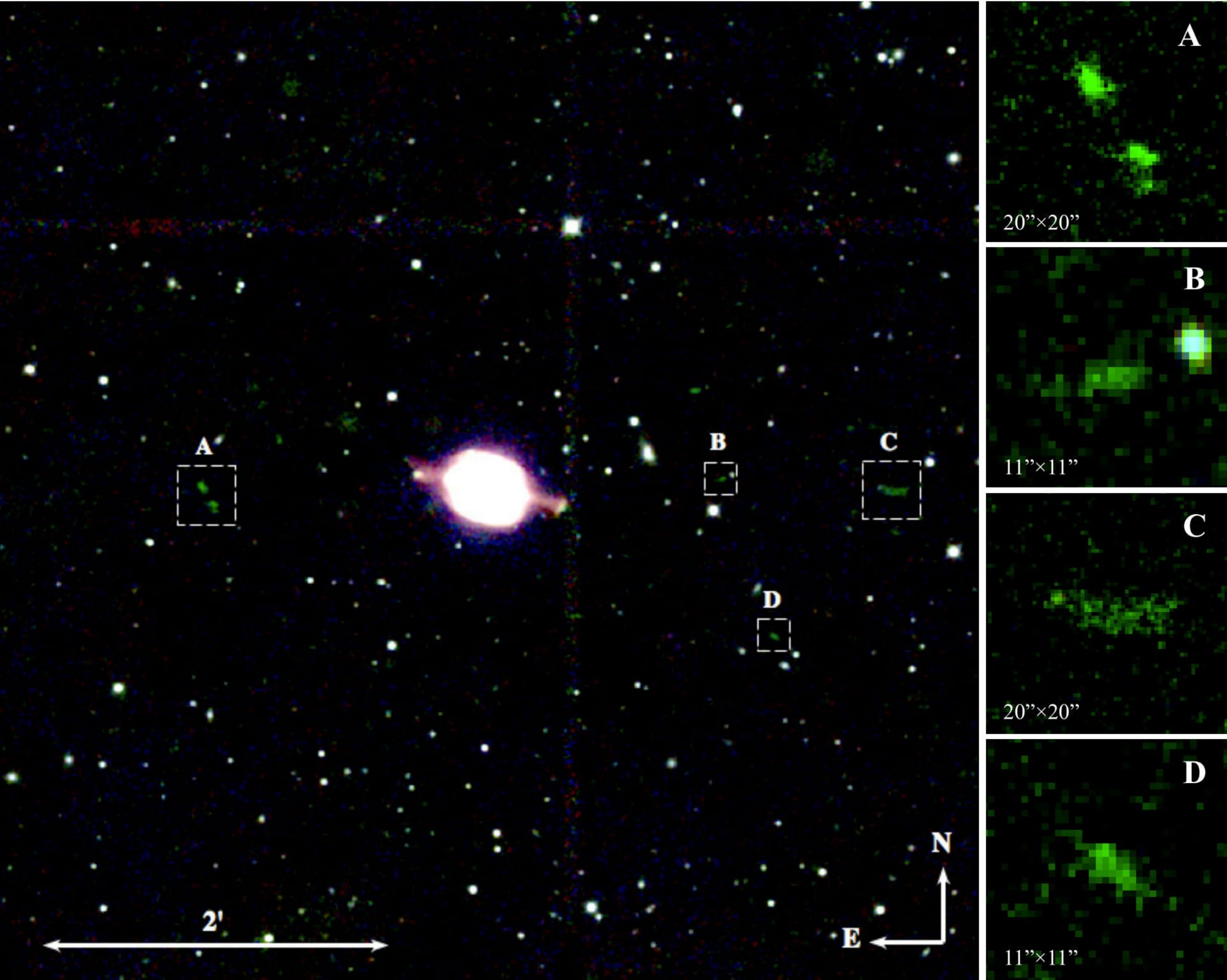}
\caption{Same as Figure~\ref{fig19} but with a much wider field 
centered on NGC\,7009.  The central nebula is saturated so that the 
faint knots/filaments in the halo are visible in H$_{2}$ (green). 
Some halo features (marked as A, B, C and D) are zoomed-in in the 
right panels.} 
\label{fig20}
\end{center}
\end{figure*}

The two well-known [N~{\sc ii}]-bright outer knots are also enhanced
in H$_{2}$ emission.  We also see faint $K_{\rm c}$ emission in 
these two knots, which might be due to H$_{2}$ 1$-$0 S(0) and/or 
H$_{2}$ 2$-$1 S(1).  The H$_{2}$ emission on the two knots is 
located further out than that of Br$\gamma$, which is better seen 
in Figure~\ref{fig19}.  A very faint, extended, [O~{\sc 
iii}]-emitting halo was first detected by \citet{MC98}; its average
surface brightness was later on measured to be $\lesssim$0.02 of 
the central bright core \citet{Corradi03}.  The eastern part of 
the halo is semi-circular, while the western halo is fragmented and
filamentary.  Through careful inspection of the CFHT images, we 
discovered in H$_{2}$ a number of knots and filamentary features 
surrounding the inner bright nebula (Figure~\ref{fig20}).  These 
condensations extend as far as $\sim$2\arcmin\ from the central 
star, and lie within the previously observed [O~{\sc iii}] halo. 
The H$_{2}$ condensations are mostly distributed along the EW 
direction.  The knots/filaments in the western halo are more 
elongated, while the eastern knots seem to be more compact.  Some 
H$_{2}$ filaments in the western halo are nearly 10\arcsec\ in 
length (Figure~\ref{fig20}).

\subsection{NGC\,7048}
\label{sec3:10}

\begin{figure*}
\begin{center}
\includegraphics[width=14.0cm,angle=0]{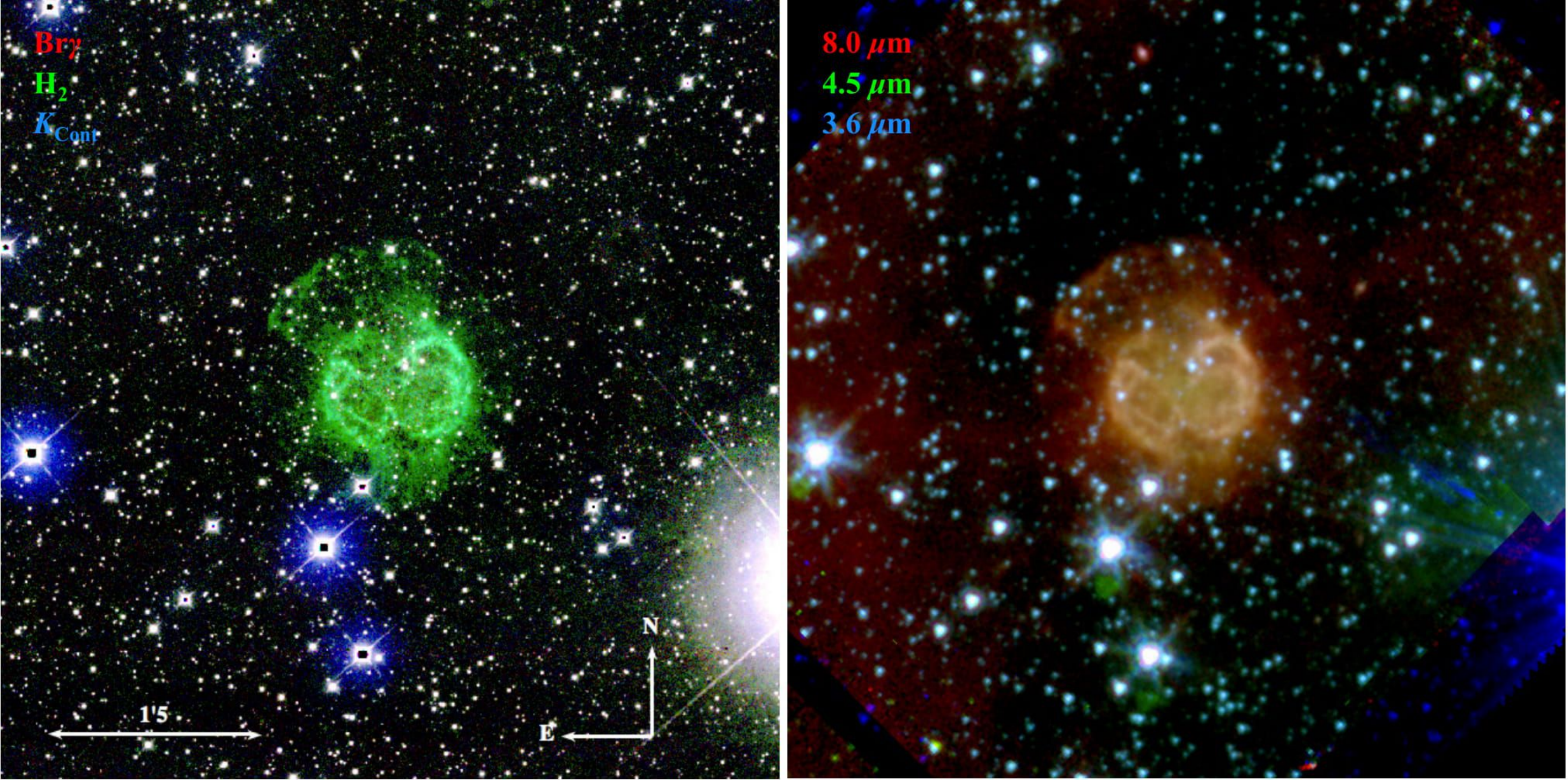}
\caption{Left: CFHT color-composite image of NGC\,7048. 
Right: \emph{Spitzer} IRAC color-composite image of NGC\,7048; 
image size and orientation are the same as left.} 
\label{fig21}
\end{center}
\end{figure*}

Despite of the brightness, the exact structure of NGC\,7048 is yet 
to be investigated in detail.  Our optical image (Figure~\ref{fig1})
shows an almost round nebula with a radius of $\sim$32\arcsec. 
There seem to be two openings oriented at PA$\sim$15$\degr$, with 
the northern opening wider than the southern one.  The [N~{\sc ii}] 
emission is enhanced at the central nebula/shell, while H$\alpha$ 
emission is more homogeneous.  There are extremely faint extensions 
along the NS direction. 
Our CFHT images reveal an extended halo dominated by the H$_{2}$ 
emission (Figures~\ref{fig2}).  The Br$\gamma$ emission is confined 
within the nebular shell.  The IRAC 8.0\,$\mu$m emission traces 
similar regions as H$_{2}$, but is more homogeneous 
(Figure~\ref{fig21}). 

\begin{figure}
\begin{center}
\includegraphics[width=7.0cm,angle=0]{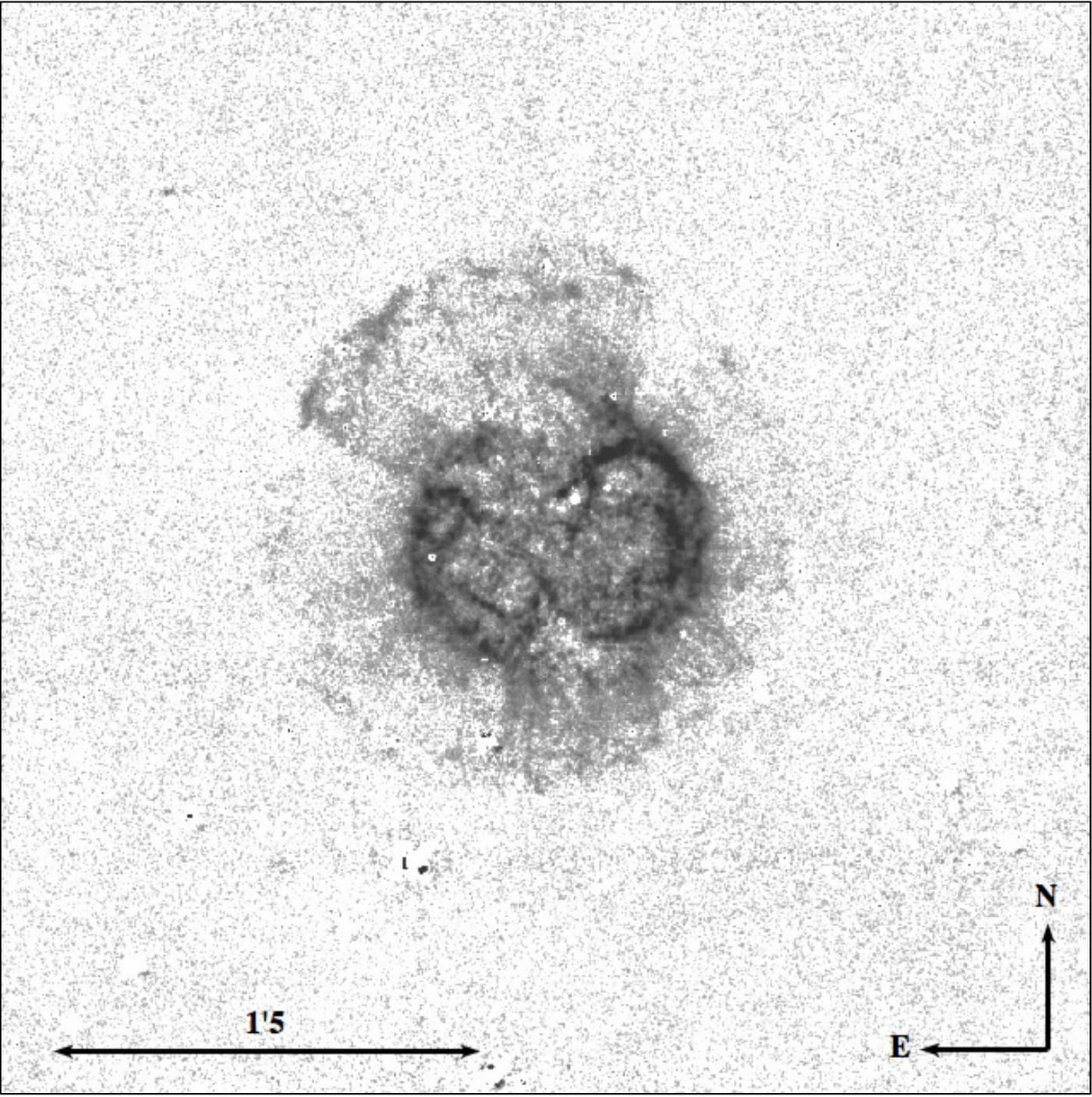}
\caption{Residual image of NGC\,7048 created by subtracting the 
scaled Br$\gamma$ image from the H$_{2}$ image.} 
\label{fig22}
\end{center}
\end{figure}

The morphology of NGC\,7048's halo is well demonstrated in the 
residual H$_{2}$ image (Figure~\ref{fig22}).  This halo has an 
overall round shape with slight deformation.  The inner 
nebula/shell is displaced southward from the geometric centre of 
the halo.  The halo emission is inhomogeneous: H$_{2}$ emission 
is in the form of radial filaments coming out from the central 
shell, which resembles what is observed in NGC\,6720.  What is 
more intriguing is that the H$_{2}$ emission is stronger in the 
north-south regions of the halo than that in the east-west.  This 
seems to be in accordance with the opening direction of the central
nebula in the optical image.  Our CFHT H$_{2}$ image also reveals 
an approximately $\infty$-shaped feature along with numerous knots  
in the nebular centre.  The shape development of the outer halo is 
probably related to the central region.  A 
detailed morpho-kinematic study of this PN is needed.

\subsection{Sh\,1-89}
\label{sec3:11}

Sh\,1-89 is a bipolar PN, as seen in deep optical images 
\citep{CS95,Manchado96,Hua97,Bohigas03}.  In the optical image, 
Sh\,1-89 comprises a seemingly edge-on waist and a pair of bipolar 
lobes oriented along PA$\sim$48$\degr$ (Figure~\ref{fig1}).  The 
two lobes are dominated by the [N~{\sc ii}] emission, which defines 
outer boundaries of the equatorial waist as well as the lobes.  The 
[O~{\sc iii}] emission mainly comes from the central region, while 
the H$\alpha$ emission is more diffuse.  In the south region of 
Sh\,1-89, $\sim$1\farcm1 from the central core region, there is a 
giant filament dominated in the [N~{\sc ii}] emission, stretching 
by nearly 2\farcm8 along the east-west direction and intersecting 
with the SW lobe.  If this filament does not belong to Sh\,1-89, it 
might be due to interaction with the ISM. 

\begin{figure}
\begin{center}
\includegraphics[width=7.0cm,angle=0]{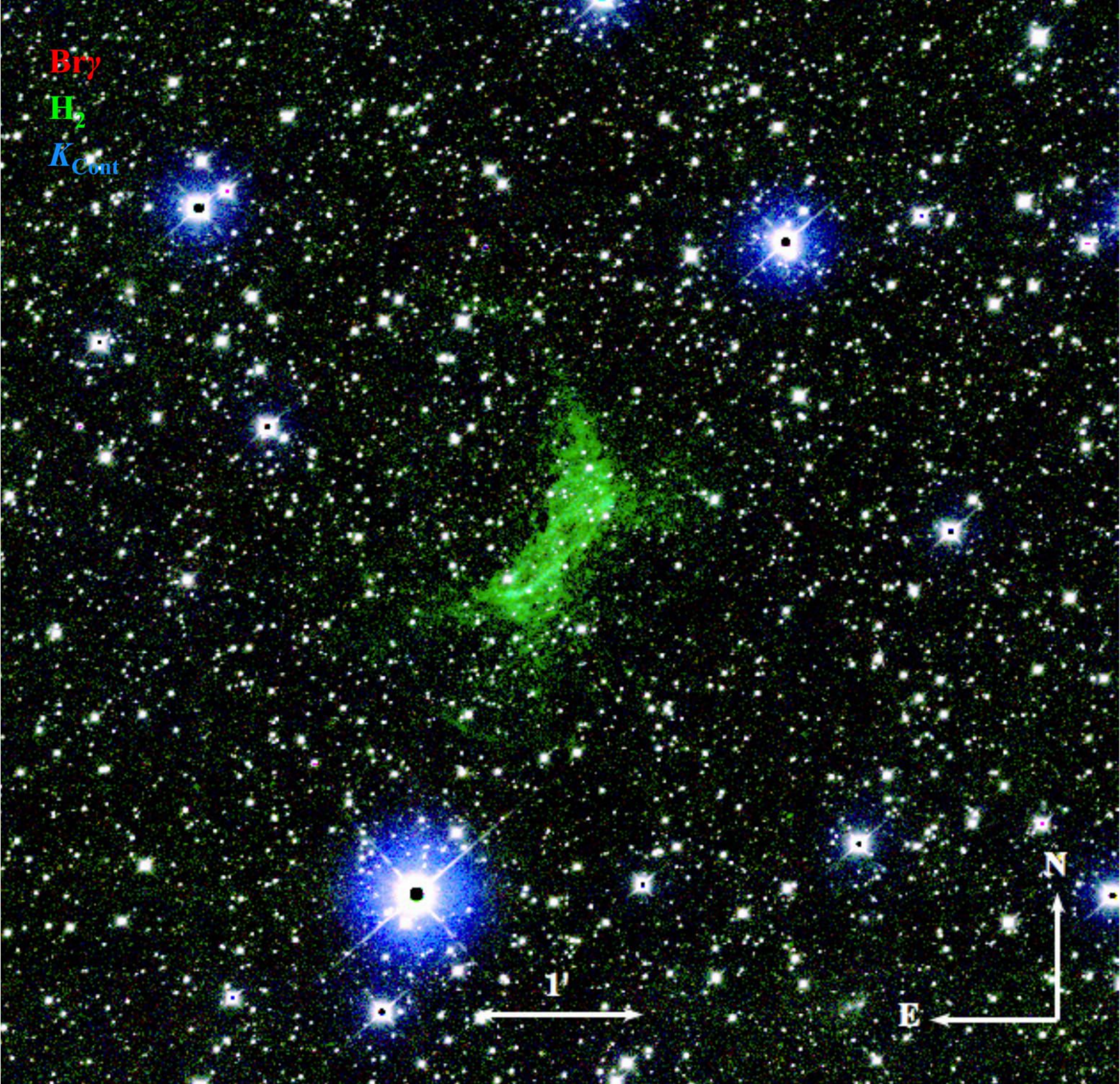}
\caption{CFHT color-composite image of Sh\,1-89.  The central torus 
and bipolar morphology is visible in H$_{2}$ emission (green).} 
\label{fig23}
\end{center}
\end{figure}

Our high-resolution near-IR images revealed very fine structures 
in the central region of Sh\,1-89 (Figure~\ref{fig2}). The central 
waist is torus-like, dominant in H$_{2}$ emission, and are 
filamentary (Figure~\ref{fig23}).  The Br$\gamma$ emission is 
faint.  The residual H$_{2}$ image clearly shows that the central 
torus of Sh\,1-89 is a tilted ring (Figure~\ref{fig24}).  This 
ring, with an angular diameter of $\sim$1\arcmin, has an apparent 
minor-to-major axis ratio of 0.189, corresponding to a inclination 
angle of $\sim$79$\degr$ with respect to the line of sight, if we 
assume that the ring is circular.  Previous H$_{2}$ imagery vaguely 
shows a tilted ring-like torus \citep{Kast96}, while our deep 
images confirm this morphology with much greater details.  There is 
extended H$_{2}$ emission beyond the central torus, which defines 
the boundaries of the bipolar lobes of Sh\,1-89.  There are also 
some ripple-like structures on both sides of the torus, along the 
NS direction, aligned with respect to the plane of the ring 
(Figure~\ref{fig24}).  These ``ripples'' in the H$_{2}$ line 
emission are probably some projected rings on the bipolar lobes of 
Sh\,1-89, a similar case to Hb\,12 \citep{KH07}.  Note that in the 
optical image, there are some diffuse ripples in the lobes in 
[N~{\sc ii}] (Figure~\ref{fig1}).

\section{Discussion} 
\label{sec4}

\subsection{The H$_{2}$ Emission and Bipolarity}
\label{sec4:1}

Detection of the molecular gas, most prominently H$_{2}$ and CO, 
toward PNe revealed that a significant fraction of nebular material
are in molecular form, and has led to new interpretations of their 
structures and evolution \citep[e.g.,][]{Isaacman84,Storey84, 
Storey87,HH86,HH89,Huggins96}.  The association between H$_{2}$ 
emission and the bipolar morphology of PNe was discussed by 
\citet{Kast94,Kast96}, who defined the so-called {\it Gatley's 
rule}, which was later confirmed by a number of imaging surveys 
\citep[e.g.,][]{HL96,Hora99,Guerrero00,Marquez13}.  It also has 
been suggested that bipolar PNe evolved from the more massive low- 
and intermediate-mass stars \citep[$\gtrsim$1.5--2.0\,$M_{\sun}$; 
e.g.,][]{Peimbert83,CS95}.  However, as revealed by sensitive 
observations, H$_{2}$ emission also exists in PNe whose morphologies
cannot be described as bipolar, but rather as ellipsoidal or 
barrel-like \citep[][]{Marquez13,Akras17}, suggesting that the 
detection of H$_{2}$ emission is not exclusive of bipolar PNe.

\begin{figure*}
\begin{center}
\includegraphics[width=14.0cm,angle=0]{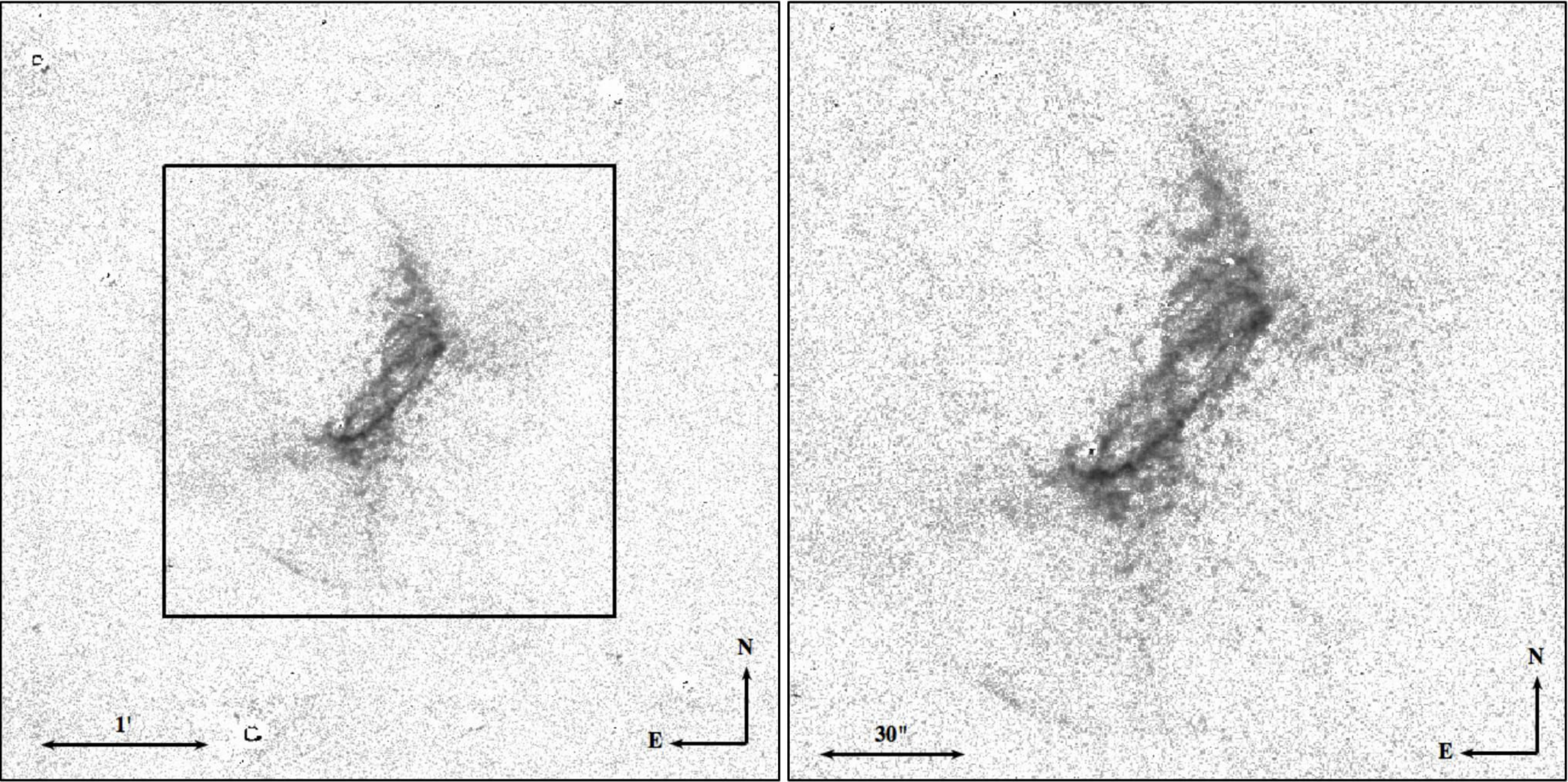}
\caption{Left: Residual image of Sh\,1-89 created by subtracting 
the scaled Br$\gamma$ image from the H$_{2}$ image.  Right: zoom-in 
of the central region as marked by a black square in the left panel,
showing details of the central ring structure.} 
\label{fig24}
\end{center}
\end{figure*}

\begin{figure*}
\begin{center}
\includegraphics[width=17.8cm,angle=0]{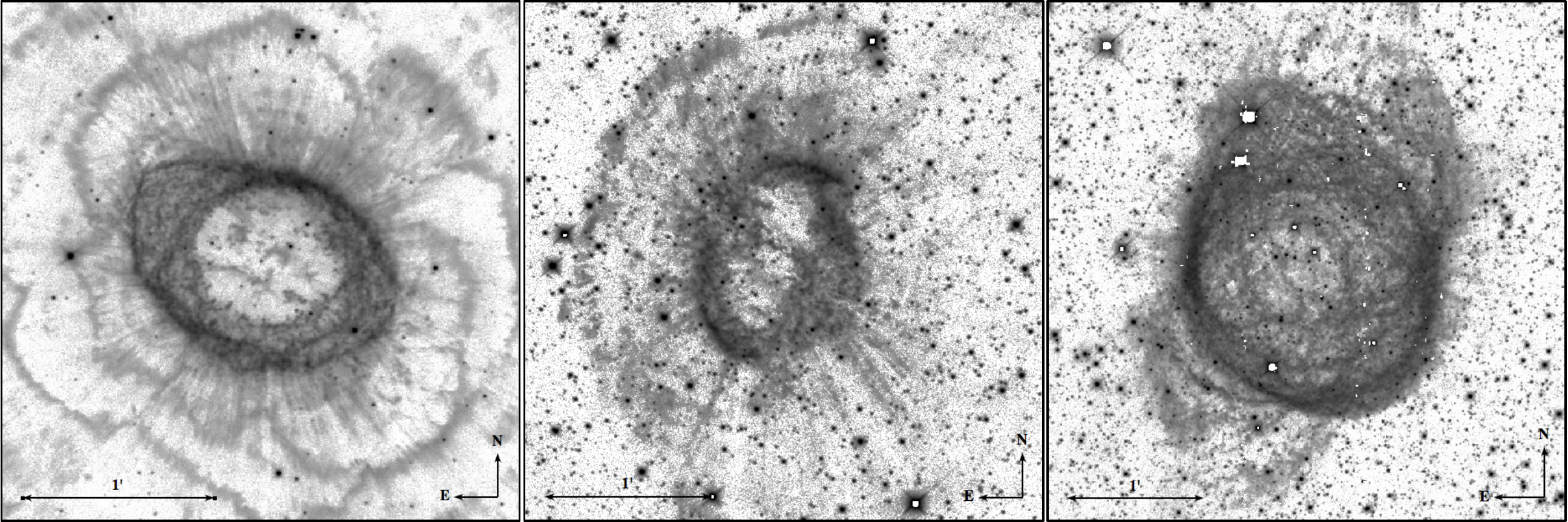}
\caption{Close-up of the CFHT H$_{2}$ images of NGC\,6720 (left), 
NGC\,6772 (middle), and NGC\,6781 (right) showing the 
fine-structure features of the central regions in details.}
\label{fig25}
\end{center}
\end{figure*}

The standard definition of ``bipolar'' is largely based on the 
apparent morphologies of PNe \citep[e.g.,][]{CS95}.  The bipolar 
PNe were grouped into two main subclasses:  those with pinched 
equatorial waists (a.k.a., ``bow-ties'' or ``hourglasses'') and 
those with broad equatorial rings (a.k.a., ``butterflies''; 
\citealt{Marquez15,Ramos17}).  These two groups of nebulae were 
previously called the early and late butterfly types, respectively,
and were even proposed to be two phases of an evolutionary sequence
\citep{Balick87}, although this postulation is disputed 
\citep{Ramos17}. 

According to that definition scheme, Hb\,12 and Sh\,1-89 are bipolar
PNe with equatorial waists, while NGC\,6445 may belong to the 
broad-ring group.  However, such definition using apparent 
morphology alone could be insufficient to account for the intrinsic
structures of all PNe.  NGC\,6720, NGC\,6772, NGC\,6781 are 
non-bipolar in appearance but clearly ring-like; they have 
been suggested to be actually bipolar nebulae viewed at large 
inclination angles \citep{Kast94}.  NGC\,7048, although not so 
obvious as those three PNe, is suggestively ring-like 
\citep{Kast96}; in our CFHT H$_{2}$ image (Figures~\ref{fig2} and 
\ref{fig22}), the central region of this PN seems to resemble a 
disrupted ring.  In the H$_{2}$ images of these four ring-like PNe,
we detected halo features surrounding the bright central nebula/ring.

The halo of NGC\,6720 has been observed from near- to far-IR 
\citep[e.g.,][]{vanHoof10}.  It is composed of an inner fragmented 
component and an outer circular one.  The structure of NGC\,6720 
is highly disputed: either a closed, ellipsoidal nebular shell 
surrounded by the remnant red giant halo \citep{Guerrero97}, 
or a bipolar PN with nearly pole-on triple biconic structure 
\citep[a counterpart of NGC\,6853,][]{Kwok08}, or a density-enhanced
main ring with bipolar lobes plus inner and outer halos 
\citep{ODell13}.  The detection of a faint crescent-shaped structure
attached to the outer circular halo of NGC\,6720 
(Figure~\ref{fig11}; similar to the case of the ``Eskimo Nebula'' 
NGC\,2392, see Figure~1 of \citealt{Zhang12b}) seems to be 
suggestive of a highly inclined bipolar nebula; however, careful 
investigation that combines the morphological and spatially-resolved
kinematic information is needed to discern the exact structure of 
this PN. 

NGC\,6781 was suggested to be a butterfly nebula, similar to 
NGC\,2346, but oriented with its polar axis close to the line of 
sight \citep{Balick87}.  It was observed in H$_{2}$ and identified 
as a low surface brightness counterpart to NGC\,6720 
\citep{Zuckerman90}.  Optical imagery of NGC\,6781 reveals a 
$\sim$1\arcmin-radius bright ring with arcs and filaments 
extending along PA$\sim$160\degr.  Morpho-kinematic observations 
of molecular lines \citep{Zuckerman90,Bachiller93,Hiriart05} and 
photoionization modeling \citep{SM06} of this PN indicate that 
the central region of NGC\,6781 probably resembles an open-ended 
cylindrical barrel oriented nearly pole-on; inside the barrel is a 
central cavity filled with fully ionized hot gas.  Recent optical 
and mid-IR observations of NGC\,6781 also suggested that it could 
be a bipolar PN oriented close to the line of sight \citep{Phil11}.
Even more recent \emph{Herschel} far-IR broad-band imaging and 
spectroscopy confirmed the nearly pole-on barrel structure of 
NGC\,6781 \citep{Ueta14}. 

Our deep H$_{2}$ imagery not only demonstrates the structure of 
NGC\,6781 in greater detail than all previous IR observations, 
but also reveals extended nebular features. 
The central H$_{2}$-bright ring and the adjacent arc features
may be morphologically explained by a highly inclined bipolar 
nebula projected on the sky.  The filaments/knots in H$_{2}$ 
emission within the elliptical inner nebular shell may be patches 
of H$_{2}$ in the central ionized region \citep{Otsuka17}. 
High-spatial resolution optical images of NGC\,6781 show knots as 
well as dark lanes across the region within the bright ring, 
indicating the existence of neutral filaments and condensations, 
which are probably associated with H$_{2}$ emission \citep{Phil11}.

The seven PNe discussed above, either bipolar or ring-like in the 
entirety of their morphologies, are mostly confined within 200~pc 
from the Galactic mid-plane (Table~\ref{targets}), which is 
consistent with the previous statistical studies 
\citep[e.g.,][]{Kast96}.

\subsection{Origin of the Molecular Structure}
\label{sec4:2}

The H$_{2}$ 2.122\,$\mu$m line emission has been used to trace 
not only the molecular component in PNe but also the interaction 
between stellar outflows and the circumstellar material.  Detailed 
imaging studies with high spatial resolution suggest that the 
bright H$_{2}$ emission from the equatorial ring of bipolar PNe, 
instead of arising from the photodissociation regions shielded 
from the bright UV radiation by the ring, actually comes from 
dense knots/clumps embedded in the ionized gas 
\citep[e.g.,][]{Cox98,Speck02,Speck03,Matsuura09,Marquez13,
Manchado15}.  Whether the H$_{2}$ in the knots within the nebular 
ring survived through the post-AGB evolution, or were destroyed and
then formed again, was discussed by \citet{vanHoof10} for the case 
of NGC\,6720.  Theoretical calculations have shown that H$_{2}$ 
may survive in the ionized region of PNe \citep{AG04}, and that 
H$_{2}$ emission in the ionized nebulae can be important, 
particularly for the PNe with high-temperature central stars 
\citep{Phillips06,AG11}.  This qualitatively explains the higher 
detection rate of H$_{2}$ emission in bipolar PNe (i.e., Gatley's 
rule), given that these PNe usually have hotter central stars. 

Several mechanisms have been proposed to explain the formation of 
these molecular knots in PNe, including the pre-existing 
high-density structures in the ISM \citep{Aluzas12} and 
fragmentation of the swept-up nebular shell after termination of 
the fast stellar wind \citep{Garcia06}.  In the latter paradigm, 
when the stellar wind declines/ceases, the hot, shocked bubble 
depressurizes, and the thermal pressure of the photoionized region
at the inner edge of the swept-up shell becomes dominant; the 
nebular shell then fragments due to the Rayleigh-Taylor 
instability or the effects of ionizing radiation in the nebula, 
leading to the formation of neutral gas clumps with comet-like 
tails (\citealt{Garcia99}; see also hydrodynamical simulations by 
\citealt{TA14,TA16} for the formation of PNe). 

As a benchmark case study of H$_{2}$ emission in PNe, the very high
spatial resolution ($\approx$60--90\,mas) imaging of the bipolar 
PN NGC\,2346 by \citet{Manchado15} demonstrates well that H$_{2}$ 
emission in the equatorial ring comes from the fragmented clumps. 
region of NGC\,2346, and appear to be uniform at lower resolutions.
In our sample, Sh\,1-89 is an archetypal counterpart to NGC\,2346 
in morphology.  In the optical, Sh\,1-89 has a bipolar structure 
(Figure~\ref{fig1}), but the H$_{2}$ emission is mostly 
concentrated in the equatorial region.  The residual H$_{2}$ image 
of Sh\,1-89 clearly shows the fine clumpy features distributed 
along a tilted equatorial ring (Figure~\ref{fig24}).  This 
similarity between the two PNe in both the macro and micro 
morphologies indicates that the formation mechanism of the H$_{2}$ 
in Sh\,1-89 might be the same as in NGC\,2346.

\begin{figure}
\begin{center}
\includegraphics[width=1.0\columnwidth,angle=0]{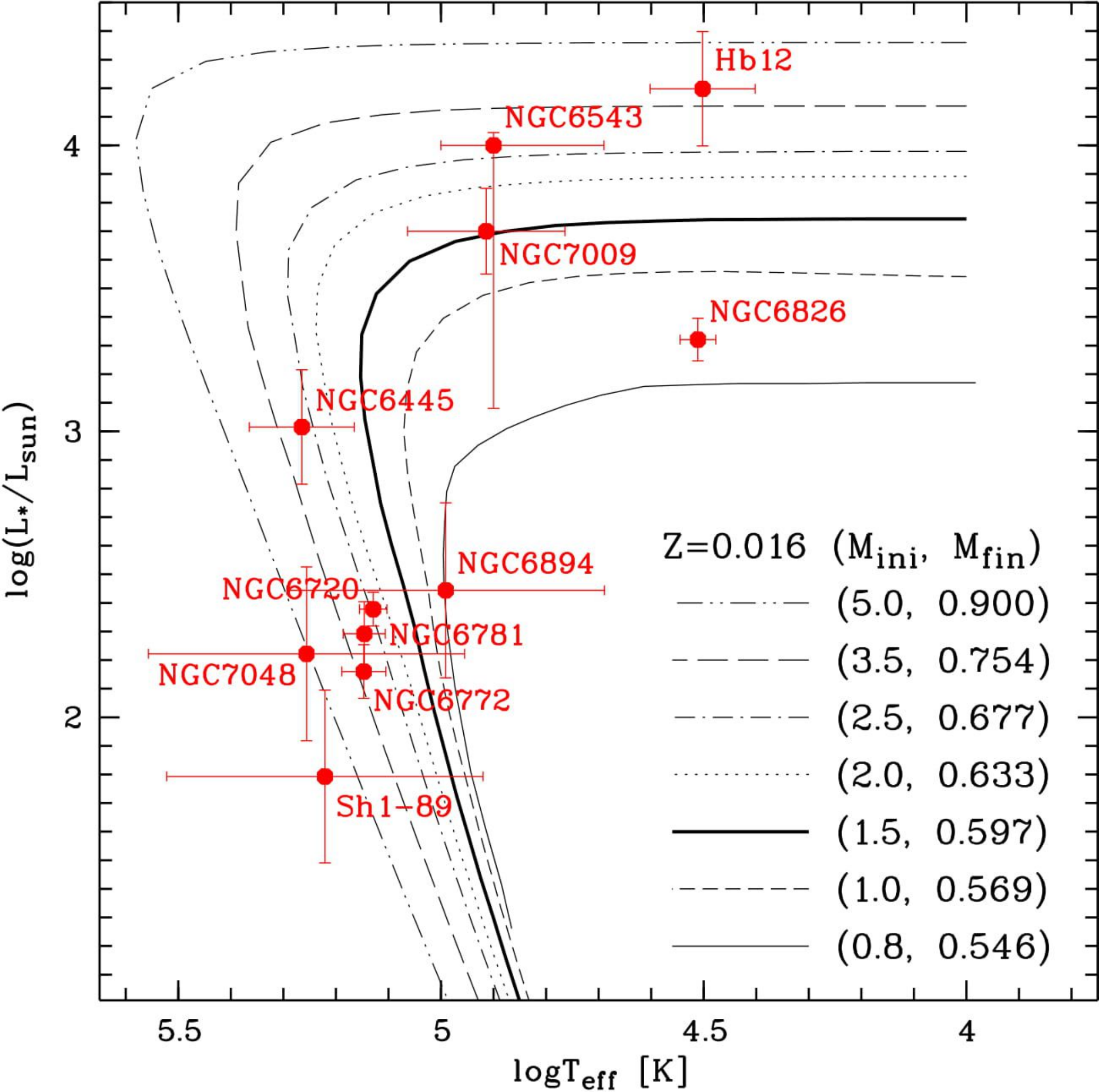}
\caption{Locations of PN central stars in the H-R diagram along 
with errors in $T_{\rm e}$ and $L_{\ast}$.  Different line types 
represent post-AGB evolutionary tracks of stars with different 
initial and final masses ($M_{\rm ini}$ and $M_{\rm fin}$, in units
of $M_{\sun}$; see the legend); model tracks are adopted from 
\citet{VW94}.} 
\label{fig26}
\end{center}
\end{figure}

Similar clumpy fine-structure features of H$_{2}$ are also 
seen in NGC\,6720, NGC\,6772 and NGC\,6781:  numerous 
H$_{2}$-emitting knots/clumps are densely located within the 
central rings of the three ring-like PNe (Figure~\ref{fig25}). 
In the H$_{2}$ images of both NGC\,6720 and NGC\,6772, we also 
observed radial filaments/rays coming out from the knots on the 
ring, which might be morphologically interpreted by the 
hydrodynamical simulations of \citet{Garcia06}:  when the effect 
of the fast stellar wind becomes negligible, the inner hot, 
shocked bubble depressurizes, and the swept-up nebular shell then 
fragments due to instability of the thermal pressure in the 
photoionized region; this process creates clumps with cometary 
tails and long, photoionized trails in between.  The fast stellar 
wind from the central star probably does not account for the 
formation of these clumps.  These radial rays are not quite 
obvious in the H$_{2}$ image of NGC\,6781, but visible in its 
residual H$_{2}$ image (see Figure~\ref{fig15}).  The fact that 
no X-ray emission was detected in these three PNe by \emph{Chandra}
\citep{Kast12} also indicates that the stellar winds from their 
central stars probably already declined. 

At the distance to NGC\,6781 \citep[0.46--0.72~kpc,][]{Frew16,
Otsuka17}, the pixel size 0\farcs3 of CFHT WIRCam corresponds to 
138--216\,AU (2.06--3.23$\times$10$^{15}$~cm), which is comparable 
to the sizes of the H$_{2}$ clumps in NGC\,2346 
\citep[112--238~AU,][]{Manchado15}.  For the more distant PN 
Sh\,1-89 (1.85~kpc), the pixel size of WIRCam corresponds to 
555\,AU (or 8.30$\times$10$^{15}$~cm).  The distance to NGC\,6772 
(1.31~kpc) is between these two PNe.  With higher spatial 
resolutions, we should expect to see the clumpy structures in the 
three PNe with much greater details. 

Excitation of H$_{2}$ emission and the nebular morphology are 
closely related to the evolutionary status of the PN central star. 
The locations of our targets in the Hertzsprung-Russell (H-R) 
diagram are shown in Figure~\ref{fig26}, where the post-AGB 
evolutionary tracks calculated by \citet{VW94} are presented. 
The effective temperatures and luminosities (as well as the 
uncertainties) of the PN central stars were adopted from the 
literature \citep{ZK93,vanHoof00,vanHoof10,Stan02,Sabbadin04,WL04,
Walsh16,Otsuka17}.  According to their locations in the H-R 
diagram, the central stars of our sample might be separated into 
two main groups:  those already on the cooling track (NGC\,6720, 
NGC\,6772, NGC\,6781, NGC\,6894, NGC\,7048, Sh\,1-89), and those 
whose temperatures are still increasing (Hb\,12, NGC\,6543, 
NGC\,6826, NGC\,7009).  This general classification of the sample 
still stands if the most up-to-date post-AGB evolutionary tracks 
of \citet{MB16} are used. 

Among the objects in the former group, NGC\,6720 is so far the best 
studied in terms of morphology and origin of H$_{2}$.  It is quite 
probable that in the post-AGB evolution, H$_{2}$ in the inner bright
ring of NGC\,6720 was first destroyed, and then formed again later 
inside the knots after the central star entered the cooling track 
\citep[see the discussion in][]{vanHoof10}.  However, if stellar 
evolution is fast enough, H$_{2}$ may still survive in the clumps. 
For the other PNe with central stars on the cooling track, H$_{2}$ 
in the equatorial ring might have formed through similar mechanism. 
The exact formation and excitation mechanisms of H$_{2}$ emission in
the halos of these PNe has never been studied, although it is the 
common sense that H$_{2}$ in a PN halo was originally formed in 
the dense AGB wind and survived.  The H$_{2}$ emission in the halos
of these PNe could be shock-excited, given that the radiation field
of the central star could be too dilute at the distance of the 
halo.  Alternatively, H$_{2}$ in the halo might have all been 
destroyed in the earlier post-AGB evolution of the central star, 
but formed later on after the halo nebular gas recombined. 


\subsection{Possible Interaction with the ISM}
\label{sec4:3}

Interaction of PNe and AGB winds (envelopes) with the ISM is 
an important process.  It can be used as a tool to 1) study the 
evolution of PNe and their halos, 2) predict the proper motion 
of the central stars of PNe, and 3) probe the structure and physical
properties of the ISM.  The interaction is also key for solving the 
``missing mass problem''.  This PN-ISM interaction usually affects 
the nebular morphology, especially the outer structures. 
Observational studies 
\citep[e.g.,][]{Borkowski90,RLP09,Sahai10,Ali12,Ali13,Zhang12a,
Sahai14,Ramos18} and theoretical modeling 
\citep[e.g.,][]{DS98,Villaver03,Villaver04,Wareing07a,Wareing07b} 
have been carried out to investigate such interaction.  The 
consequences of interaction with the ISM are visually manifested as 
displacement of the central star from the geometric centre of the 
nebula, or shifting of the nebular shell from the outer halo.  The 
former case is usually seen in the optical bands, while the latter 
can be witnessed in the IR \citep[e.g.,][]{RLP09}.  \citet{Ali12} 
presented a catalogue of 117 Galactic PNe that have interaction with 
the ISM, thus providing a unique database for such kind of study.

\begin{figure}
\begin{center}
\includegraphics[width=1.0\columnwidth,angle=0]{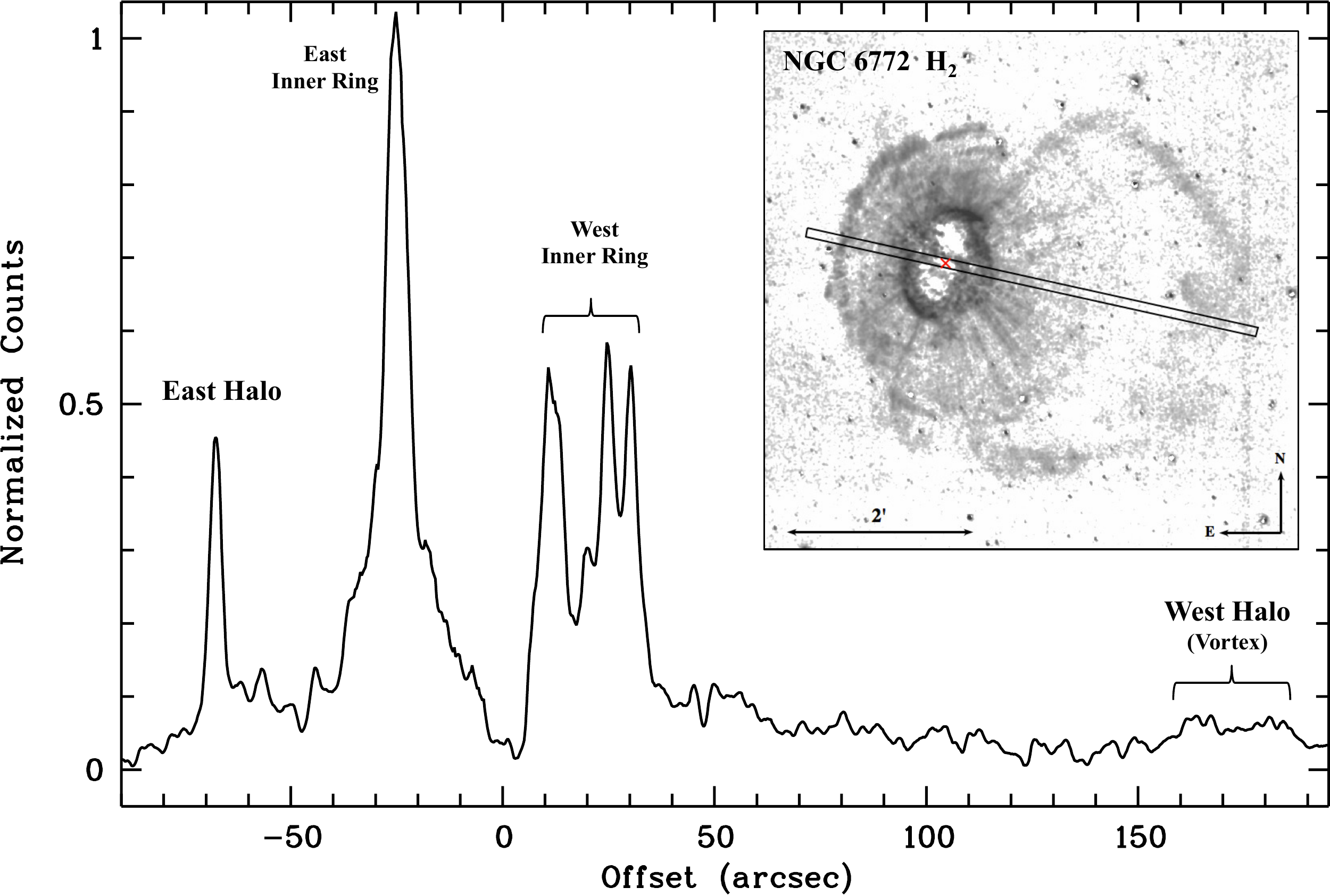}
\caption{H$_{2}$ emission profile along a cut through the central 
star of NGC\,6772 at PA=77$\degr$ (see the inset image; the cut 
width is 20 pixels, corresponding to 6\arcsec), in parallel with 
the possible proper motion direction (Figure~\ref{fig13}, right). 
Both the image and emission profile have been slightly smoothed to 
reduce noise.  Horizontal axis is the distance (in arcsec) from 
the central star (red cross marked in the inset) defined to 
increase from east to west.} 
\label{fig27}
\end{center}
\end{figure}

\begin{figure*}
\begin{center}
\includegraphics[width=17.8cm,angle=0]{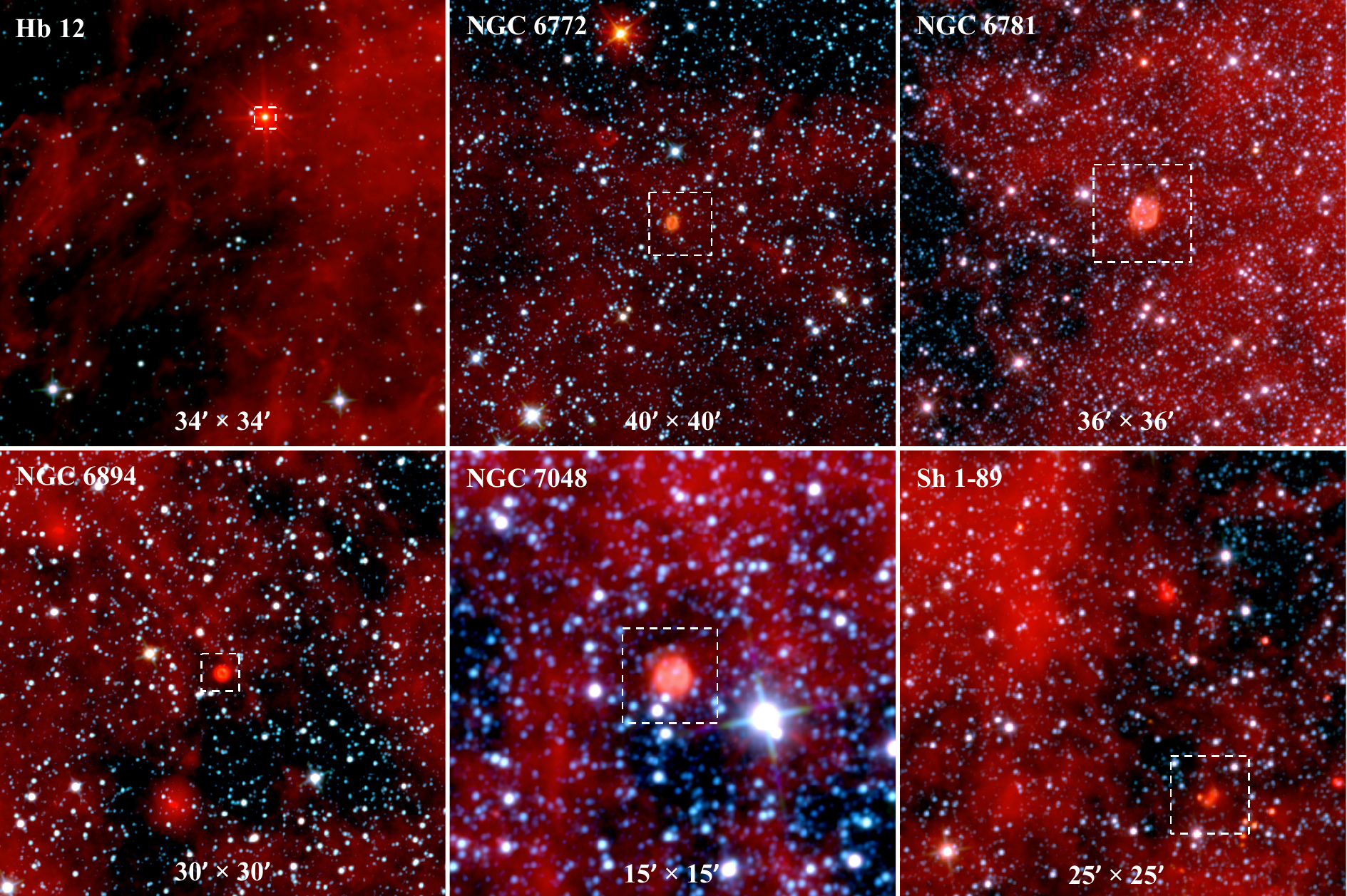}
\caption{\emph{WISE} color-composite pictures of Hb\,12, NGC\,6772,
NGC\,6781, NGC\,6894, NGC\,7048 and Sh\,1-89 created with the $W$1 
(3.4\,$\mu$m, blue), $W$2 (4.6\,$\mu$m, green) and $W$3 
(12\,$\mu$m, red) bands.  Image size is indicated at the bottom of 
each panel; north is up and east to the left.  In each panel, the 
white dashed square marks the field of view of the CFHT images of 
this PN in Figure~\ref{fig2}.} 
\label{fig28}
\end{center}
\end{figure*}

NGC\,6772 is the best candidate in our sample that shows possible 
interaction with the ISM.  An extended, asymmetric halo is seen in 
H$_{2}$ emission (Figure~\ref{fig12}, left).  The inner nebular 
shell, bright in the optical, is shifted towards the east.  The 
eastern part of its halo seems to be compressed and limb-brightened,
while the western halo is more extended in the form of a giant 
``loop''.  The eastern halo is brighter than the distorted 
western part in the mid-IR \citep{RLP09}.  This difference in 
brightness is also seen in our H$_{2}$ image; moreover, we notice 
that the eastern halo shell is somewhat fragmented. This morphology
indicates that NGC\,6772 is probably interacting with the ISM. 

We made a radial cut through the central star of NGC\,6772 to 
check the emission profile of H$_{2}$, as shown in 
Figure~\ref{fig27}, where the effect of enhanced shell-brightening 
due to interaction with the ISM is seen in the eastern halo (while 
the emission of the faint, much extended, diffuse western halo is 
marginally visible).  The inner ring of NGC\,6772 also seems to be 
distorted: emission of the west part seems to be more diffuse than 
the east, suggesting that interaction with the ISM not only shapes 
(and fragments) the outer halo of NGC\,6772, but might also 
deform its inner region.  Compared to H$_{2}$, Br$\gamma$ emission 
in NGC\,6772 is more homogeneous and only confined within the inner
region (see Figure~\ref{fig2}), thus subtraction of the Br$\gamma$
image from the H$_{2}$ image probably does not affect H$_{2}$ 
emission in the halo.  Hence, the emission profile of H$_{2}$ in 
NGC\,6772, at least for the outer halo (i.e., the East and West 
Halos defined in Figure~\ref{fig13}), is reliable. 

An asymmetric halo seems to be surrounding the central nebula of 
NGC\,7048, with H$_{2}$ emission enhancement generally along 
the north-south direction (Figures~\ref{fig2} and \ref{fig22}). 
We anticipate there might be interaction between NGC\,7048 and the 
ISM.  The halo features in NGC\,6781 are very faint in H$_{2}$ 
emission, compared to the other ring-like objects such as NGC\,6720,
NGC\,6772, and NGC\,7048 (see description in Section~\ref{sec3:6}).
However, the \emph{Herschel} Planetary Nebula Survey (HerPlaNS) 
found a surrounding halo in NGC\,6781 in far-IR emission 
\citep[through the PACS/SPIRE broad-band imaging;][]{Ueta14}, 
indicating the existence of a cold dust component.  No halo is seen
in NGC\,6894 in H$_{2}$, although this PN is ring like.  It has 
been proposed  that the halo of NGC\,6894 was stripped by the ISM 
\citep{SZ97}. 

We checked the surroundings of our targets to seek the 
possibility of PN-ISM interaction using the \emph{WISE} archival 
images obtained from IRSA (see Section~\ref{sec2:5}).  The 
\emph{WISE} images show that the four ring-like objects, NGC\,6772,
NGC\,6781, NGC\,6894 and NGC\,7048, as well as two bipolar nebulae,
Hb\,12 and Sh1-89, might be located within huge volumes of the ISM
that are seen in 12\,$\mu$m emission (Figure~\ref{fig28}); however,
whether these ISM are truly associated (i.e., interacting) with 
these PNe, or simply the foreground/background material projected 
on the sky, needs to be confirmed.  In our optical image, a giant 
[N~{\sc ii}]-emitting filament intersects with the southwest lobe 
of Sh\,1-89 (Figure~\ref{fig1}), which might be indicative of 
interaction with the ISM.  To confirm this, a follow-up study of 
Sh\,1-89 through deep, wide-field optical imaging and 
high-dispersion spectroscopy will be underway (X.\ Fang et al.\ 
2018, in preparation). 

In the \emph{WISE} image, we noticed a long feature close to 
NGC\,6894 in 12\,$\mu$m emission, stretching along the NE-SW 
direction (Figure~\ref{fig28}, bottom-left).  This feature 
generally coincides in spatial distribution with the ``stripes'' 
seen in the H$\alpha$ image \citep[][Figure~1 therein]{SZ97}. 
Based on the orientation and morphologies seen in H$\alpha$, 
\citet{SZ97} suggest that these ``stripes'' near (more 
specifically, to the northwest of) NGC\,6894 may actually belong 
to the ionized halo of this PN that was stripped by the ISM. 
However, the detailed stripe-like structures as shown in the 
H$\alpha$ image cannot be resolved in Figure~\ref{fig28} due to 
low resolution (6\farcs1--6\farcs5) of the \emph{WISE} image. 

We also checked the \emph{WISE} archive for the other five 
PNe (NGC\,6445, NGC\,6543, NGC\,6720, NGC\,6826, and NGC\,7009) 
in our sample, and found that either they are so bright in the 
\emph{WISE} bands (especially in $W$3) that they outshines the 
nearby sky regions, or there is no surrounding ISM nearby.  It is 
noteworthy that NGC\,6772, NGC\,6781 and NGC\,6894 have already 
been identified as interacting PNe in 
\citet{Ali12}.\footnote{NGC\,6826 was also considered as an 
interacting PN in \citet{Ali12}, but it is too bright in the 
\emph{WISE} bands and no obvious ambient ISM is seen nearby.} 

As aforementioned, interaction with the ISM can strongly 
affects the outer structure of a PN \citep[e.g.,][]{Wareing07b}. 
The nebula is compressed and limb-brightened in the direction of 
motion when a PN is moving through the ISM. 
In the residual H$_{2}$ image of NGC\,6772 (Figure~\ref{fig13}), 
a vortex structure is clearly seen at the westernmost part of its 
outer halo, extending to $\sim$3\arcmin\ from the nebula center. 
This vortex, uncovered by the \emph{Spitzer} image 
(Figure~\ref{fig12}, right), is now discovered for the first time 
in this PN, and has an angular extent of $\sim$2\arcmin, which 
corresponds to 0.75~pc \citep[at a distance of 
1.3~kpc,][]{Stan08,Frew16}.  This physical size of vortex is 
consistent with the hydrodynamical simulations of 
\citet{Wareing07a}, who predicted the size scale of vortices to 
be in the range 0.1--10~pc.  A close inspection of 
Figure~\ref{fig13} also reveals that the AGB halo of NGC\,6772 
seems to have multiple layers: the western part of the outermost 
layer is in the process of disruption, while the eastern part is 
compressed, probably due to interaction with the ISM.

\subsection{Mass-loss History: A Case Study of Hb\,12}
\label{sec4:4} 

Extended structures of PNe are a result of complex mass-loss
process combined with stellar wind interaction (sometimes, also 
interaction with the ISM).  A careful morphological study of PNe 
with an aid of morpho-kinematic modeling can help us to better 
understand not only the wind interaction but also the mass-loss 
history of their immediate progenitor, the AGB stars.  Here we 
briefly introduce a case study of Hb\,12, which, compared to the 
other two bipolar PNe (NGC\,6445 and Sh\,1-89) in our sample, has 
much better defined bipolar lobes as well as micro-structures. It 
is thus relatively easier to construct a morphological model for 
Hb\,12. 

The three coaxial, nested bipolar lobes observed in Hb\,12 probably
indicate that the collimated bipolar outflows (jets) were ejected 
at different episodes.  This seems to be, to some extent, 
similar to the case of M\,2-9, which probably have experienced 
multiple mass-loss events due to interaction of the binary stellar 
system at the center \citep{Castro12}. Photometric observations 
suggest that the central star of Hb\,12 could be a close binary 
system, which may be responsible for its bipolar morphology 
\citep{Hsia06}.  We thus speculate that the multiple co-axial 
bipolar lobes of Hb\,12 reflect the mass-loss events that may also 
be related to the binary central star.  However, very high 
resolution (such as the interferometric) observations are needed 
to resolve the physical structure and kinematics of the central 
equatorial region of Hb\,12, which may in turn tell us the true 
nature of the central star. 

\begin{figure}
\begin{center}
\includegraphics[width=1.0\columnwidth,angle=0]{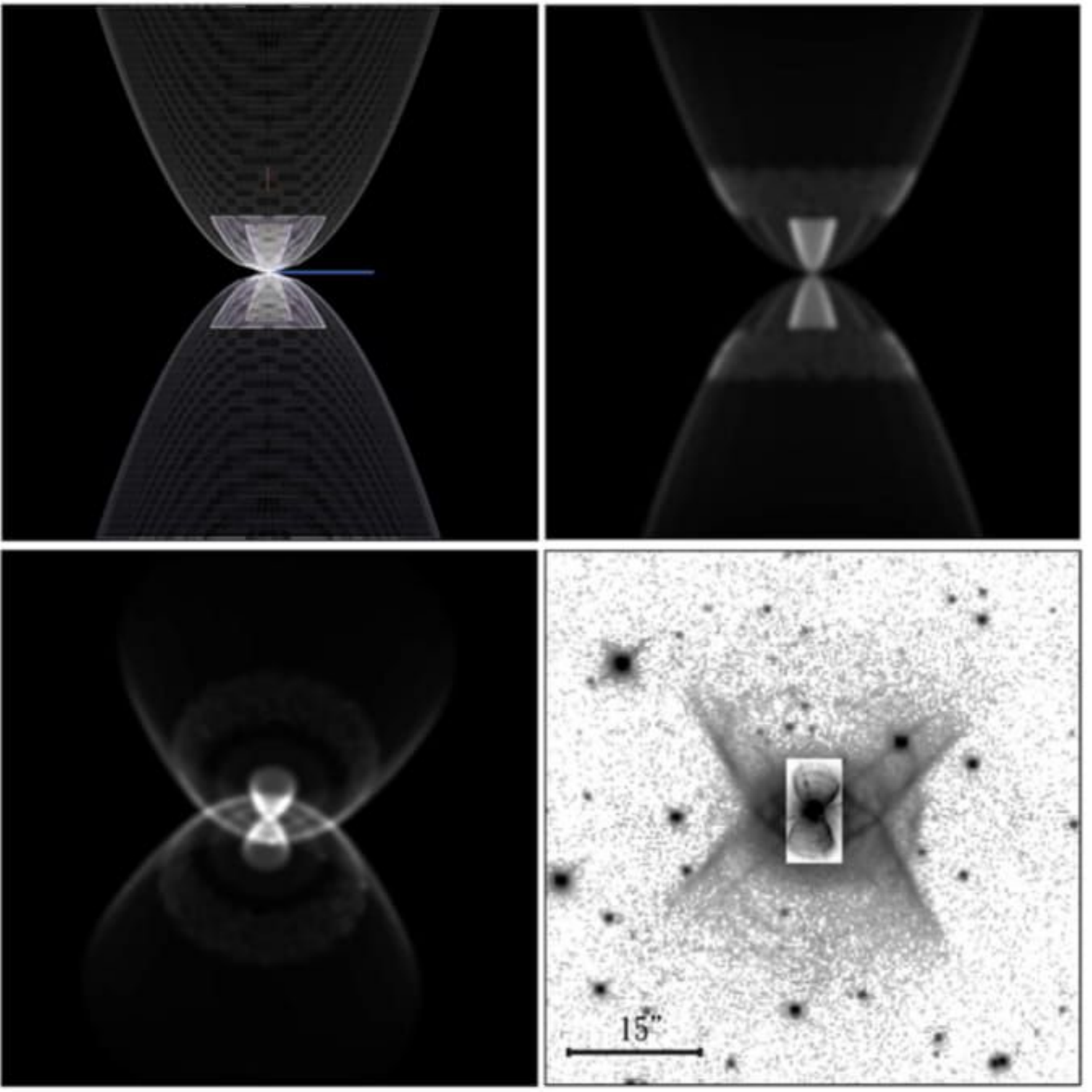}
\caption{Comparison between the 3D {\sc shape} model (top and
bottom-left panels) and the optical+IR image (bottom-right) of
Hb\,12.  Top-left panel shows the model with an inclination angle 
of 0$\degr$ (i.e., the polar axis lies in the plane of the sky), 
and the top-right panel is the rendered image with the same 
inclination angle.  Bottom-left panel is the rendered image tilted 
with an inclination angle of $\sim$45$\degr$ (and PA = 2$\degr$) 
to reproduce the equatorial ``eye'' structures seen in the 
optical$+$IR image presented in the bottom-right panel, which is 
displayed by our CFHT WIRCam H$_{2}$ image and the inset \emph{HST} 
H$\alpha$ image at the centre.} 
\label{fig29}
\end{center}
\end{figure}

The two pairs of knots along the direction of the polar axis 
signifies interaction of the jets with the circumstellar material. 
In order to replicate the main features in Hb\,12 as observed in 
our deep near-IR images (see the description in 
Section~\ref{sec3:1}), we tried to construct a 3D morphological 
model using the software {\sc shape} \citep{Steffen06,Steffen11}. 
Figure~\ref{fig29} shows a comparison of the 3D model with the 
H$_{2}$+[N~{\sc ii}] image of Hb\,12 (see also Figure~\ref{fig4}, 
left); detailed description of the model will be presented in a 
subsequent paper (C.-H.\ Hsia et al.\ 2018, in preparation). 
Although velocity information is still needed to better model the 
3D structures, at this stage we may conclude that it is 
possible to reproduce the central eye-shaped structure without an 
equatorial torus.

\section{Summary and Conclusion} 
\label{sec5} 

We report sensitive, near-IR observations of a sample of 
Galactic PNe using WIRCam on CFHT, aiming to search extended 
nebular structures through deep imaging in the H$_{2}$ 
2.122\,$\mu$m line emission. 
Thanks to the large field of view and high dynamic range of the 
detector, we observed highly intriguing nebular structures in a 
few objects.  In several PNe of our sample, we detected extended 
structures in H$_{2}$ emission, and some of these structures were 
detected for the first time.  These CFHT images were studied 
in contrast with the available optical images obtained from the 
\emph{HST} archive and ground-based telescopes.  The morphologies 
of our targets in H$_{2}$ were described in detail and their 
significance in the shaping of PNe, stellar wind interaction, AGB 
mass loss, and the PN-ISM interaction, is discussed. 

A particular case is the bipolar nebula Hb\,12.  In the CFHT 
H$_{2}$ image of Hb\,12 we observed 1) three co-axial, nested 
bipolar lobes, 2) a few arc features on either side of the central 
star probably due to inclined co-axial rings aligned along the 
outer bipolar lobes, and 3) two pairs of faint, point-symmetric 
knots about the central star distributed along the polar axis with 
slight misalignment. The outer pair of knots is the first detection
in Hb\,12.  The overall morphology of Hb\,12 resembles several 
other hourglass-shaped nebulae, including the ``Minkowski's 
Butterfly'' M\,2-9, which has a binary stellar system at the center
and may have experienced multiple episodes of mass loss.  We 
briefly introduced a morphological {\sc shape} model of this PN, 
which will be described in detail in a separate paper. 

Structures of our sample PNe imaged in the H$_{2}$ line 
emission were discussed in the context of the long-established 
H$_{2}$ emission versus bipolarity relation (i.e., {\it Gatley's 
rule}).  The ring-like objects in our sample might actually be the 
bipolar nebulae highly inclined with respect to the line of sight. 
However, this is merely a speculation, for these ring-like nebulae,
based on the morphologies as seen in H$_{2}$; careful investigation
utilizing kinematic information is needed.  For several ring-like 
PNe that were previously suggested to be bipolar, our high-spatial 
resolution imaging in H$_{2}$ resolves their central regions into 
numerous knotty/filamentary fine-structure features, which confirms
the previous finding that H$_{2}$ emission mostly comes from the 
discrete knots/clumps embedded within the fully ionized gas at the 
equatorial regions of PNe. 

For the first time, the asymmetric halo of NGC\,6772 was 
fully unveiled in H$_{2}$ emission, which probably is a result of 
interaction with the ISM.  At the westernmost region of the highly 
distorted halo of NGC\,6772, we spotted a ``vortex'', which could 
be due to the hydrodynamical effects.  Several other objects in 
our CFHT sample have also been reported to be interacting PNe.  We 
checked the \emph{WISE} database, and found that more than half of 
our sample, including the two bipolar nebulae Hb\,12 and Sh\,1-89, 
seem to be surrounded by huge volumes of the ISM visible in 
12\,$\mu$m.  Whether these ISM are truly associated with the PNe, 
or are simply the foreground/background projections, is still 
unclear.  It is worthwhile to investigate these PNe with respect 
to their surrounding environments.  The southwest lobe of the 
butterfly-shaped Sh\,1-89 is intersected by a giant filament that 
is bright in [N~{\sc ii}], which might also be indicative of 
interaction with the ISM.  Besides, in the residual H$_{2}$ image, 
the central equatorial region of Sh\,1-89 clearly displays a 
tilted ring structure. 

We observed a giant, patchy halo in H$_{2}$ surrounding the bright
central region of NGC\,6543; the overall morphology of these H$_{2}$
features generally follows the optical features in [S~{\sc ii}] and
[N~{\sc ii}].  In NGC\,7009, we found very faint, knotty/filamentary
halo features in H$_{2}$ that are distributed as far as 
$\sim$2\arcmin\ from the central star.  These halo H$_{2}$ features
might be shock excited by the fast stellar wind that escaped from 
the central nebula; this possible mechanism seems to be consistent 
with the \emph{Chandra} detection of extended X-ray emission within
the inner hot bubbles of the two PNe. 

This work well demonstrates the advantage of IR, wide-field 
sensitive imaging in studying the extended halo structures of PNe 
that are usually invisible in the optical bands.  Given its 
extensive distribution, H$_{2}$ is an excellent tracer of both the 
ionized and molecular regions of a PN.  Multi-wavelength 
observations are essential to search the full extent of PNe; 
moreover, kinematic measurements is needed to to investigate the 
intrinsic structures of PNe.  Despite our detailed description of 
the CFHT near-IR images, interpretation of some of the nebular 
structures presented in this work is still speculative, due to the 
lack of kinematic information.  Follow-up morpho-kinematic studies 
of some of these PNe is highly desirable. 


\acknowledgments

Part of the data presented here were obtained with ALFOSC, which 
is provided by the Instituto de Astrof\'{i}sica de Andaluc\'{i}a 
(IAA-CSIC) under a joint agreement with the University of Copenhagen
and NOSTA, and the 1.5\,m telescope at San Pedro M\'{a}rtir of the 
National Astronomical Observatory (OAN) operated by Universidad 
Nacional Aut\'{o}noma de M\'{e}xico (UNAM). 
This work was supported by the Research Grants Council of Hong Kong 
Special Administrative Region, China (project number HKU7031/10P 
and HKU7062/13P). C.-H.~H.\ acknowledges financial support from the 
Science and Technology Development Fund of Macau (project No.\ 
119/2017/A3 and 061/2017/A2).  M.A.G.\ acknowledges support of the 
grant AYA~2014-57280-P, co-funded with FEDER funds.  This research 
uses data obtained through the Telescope Access Program (TAP), which
has been funded by the National Astronomical Observatories of China, 
the Chinese Academy of Sciences (the Strategic Priority Research 
Program ``The Emergence of Cosmological Structures'' Grant No.\
XDB09000000), and the Special Fund for Astronomy from the Ministry 
of Finance.  This research made use of the SIMBAD database, operated
at CDS, Strasbourg, France, and of NASA’s Astrophysics Data System 
Bibliographic Services.  We thank Quentin A.\ Parker and Foteini 
Lykou for reading of the manuscript and comments.  This research 
has made use of the \emph{HST} archival data from MAST, the Mikulski
Archive for Space Telescope at the Space Telescope Science Institute
(STScI), which is operated by the Association of Universities for 
Research in Astronomy, Inc., under NASA contract NAS5-26555. 
Support for MAST for non-\emph{HST} data is provided by the NASA 
Office of Space Science via grant NAG5-7584 and by other grants and
contracts.  This publication makes use of data products from the 
\emph{Wide-field Infrared Survey Explorer} (\emph{WISE}), which is 
a joint project of the University of California, Los Angeles, and 
the Jet Propulsion Laboratory/California Institute of Technology, 
funded by the National Aeronautics and Space Administration. 
This research also utilized the software SAOImage~DS9 \citep{JM03} 
developed by Smithsonian Astrophysical Observatory. 

\vspace{5mm}

\emph{Facilities:} CFHT (WIRCam), NOT (ALFOSC), OAN-SPM:1.5\,m


\end{document}